\documentclass[apj]{emulateapj}

\usepackage{times}

\shorttitle{A Systematic Survey of High Temperature Emission}
\shortauthors{Warren, Winebarger, \& Brooks}

\begin{document}

%% ------------------------------------------------------------------------------------------
%% --- TITLE PAGE ---------------------------------------------------------------------------
%% ------------------------------------------------------------------------------------------

\title{A Systematic Survey of High Temperature Emission in Solar
  Active Regions} 

\author{Harry P. Warren\altaffilmark{1}, Amy R. Winebarger\altaffilmark{2}, and David
  H. Brooks\altaffilmark{3}}
\altaffiltext{1}{Space Science Division, Naval Research Laboratory, Washington, DC 20375}
\altaffiltext{2}{NASA Marshall Space Flight Center, VP 62, Huntsville, AL 35812}
\altaffiltext{3}{College of Science, George Mason University, 4400 University Drive,
  Fairfax, VA 22030}

%% ------------------------------------------------------------------------------------------
%% --- ABSTRACT -----------------------------------------------------------------------------
%% ------------------------------------------------------------------------------------------

\begin{abstract}
  The recent analysis of observations taken with the EIS instrument on \textit{Hinode}
  suggests that well constrained measurements of the temperature distribution in solar
  active regions can finally be made. Such measurements are critical for constraining
  theories of coronal heating. Past analysis, however, has suffered from limited sample
  sizes and large uncertainties at temperatures between 5 and 10\,MK. Here we present a
  systematic study of the differential emission measure distribution in 15 active region
  cores. We focus on measurements in the ``inter-moss'' region, that is, the region
  between the loop footpoints, where the observations are easier to interpret. To reduce
  the uncertainties at the highest temperatures we present a new method for isolating the
  \ion{Fe}{18} emission in the AIA/\textit{SDO} 94\,\AA\ channel. The resulting
  differential emission measure distributions confirm our previous analysis showing that
  the temperature distribution in an active region core is often strongly peaked near
  4\,MK. We characterize the properties of the emission distribution as a function of the
  total unsigned magnetic flux. We find that the amount of high temperature emission in
  the active region core is correlated with the total unsigned magnetic flux, while the
  emission at lower temperatures, in contrast, is inversely related. These results provide
  compelling evidence that high temperature active region emission is often close to
  equilibrium, although weaker active regions may be dominated by evolving million degree
  loops in the core.
\end{abstract}


%% ------------------------------------------------------------------------------------------
%% --- INTRODUCTION -------------------------------------------------------------------------
%% ------------------------------------------------------------------------------------------

\section{Introduction}

The distribution of temperatures in the solar atmosphere holds many important clues as to
how the solar corona is heated. Coronal loops observed at temperatures near 1\,MK, for
example, often have very narrow temperature distributions
\citep{aschwanden2005,tripathi2009,warren2008} and are evolving
\citep[e.g.,][]{winebarger2003,ugarteurra2009,tripathi2010,mulumoore2011}, suggesting that
these loops are far from equilibrium. Coronal emission at higher temperatures
($\sim$4\,MK) appears to behave differently. There is some evidence that the high
temperature emission in the core of an active region is close to equilibrium
\citep{winebarger2011,warren2011}, suggesting that heating events must occur at high
frequency to prevent loops from cooling.

This difference in behavior between loops at different temperatures appears puzzling, but
may be explained by recent work on wave models of coronal heating.
\cite{vanballegooijen2011} and \cite{asgari-targhi2012} have studied the dissipation of
Alfv{\'e}n waves in the chromosphere and corona. The heating rate that they derive is
highly localized at the loop footpoint and heating events occur at high frequency. This
implies that short loops that are strongly heated will have high apex temperatures and,
because heating events occur frequently, they will be close to equilibrium. For longer
loops that are heated more weakly the apex temperature will be lower. For such loops it is
possible that no equilibrium exists regardless of the frequency of heating events
\citep[e.g.,][]{serio1981,peter2012}. This would give rise to evolving loops at lower
temperatures.

The observational evidence for equilibrium loops at high temperatures, however, is
limited. \cite{winebarger2011} and \cite{warren2011} presented the emission measure
analysis for small areas in two active regions. In their analysis they find emission
measure distributions that are strongly peaked, suggesting that loops are not evolving
through a broad range of temperatures. Because of the limited sample size, it is unclear
how general these results are. \cite{tripathi2011}, for example, have found somewhat
broader emission measure distributions for two other active regions. \cite{viall2011} have
analyzed the temporal evolution of the emission in yet another active region and find
evidence for evolving loops, even in the core.

\begin{figure*}[t!]
\centerline{\includegraphics[clip,angle=90,scale=0.67]{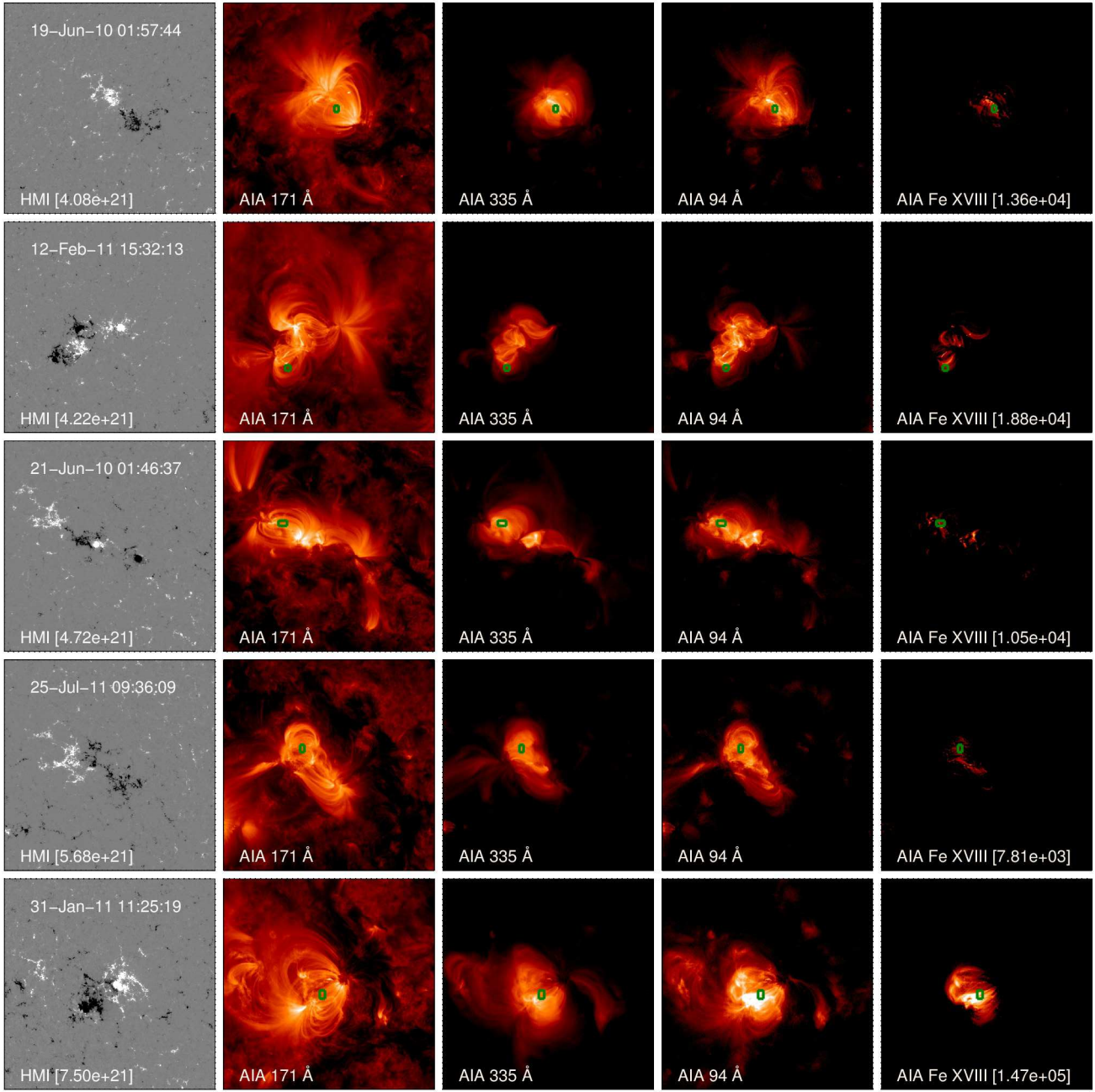}}
\caption{AIA and HMI observations of solar active regions. The regions are presented in
  order of increasing total unsigned magnetic flux. Every image at a particular wavelength
  is displayed with the same scaling. The green boxes represent the regions selected to
  compute the emission measure distribution. The numbers in brackets are the fluxes given
  in Table~\protect{\ref{table:list}}. Data for regions 1--5 are shown here.}
\label{fig:summary1}
\end{figure*}

\begin{figure*}[t!]
\centerline{\includegraphics[clip,angle=90,scale=0.67]{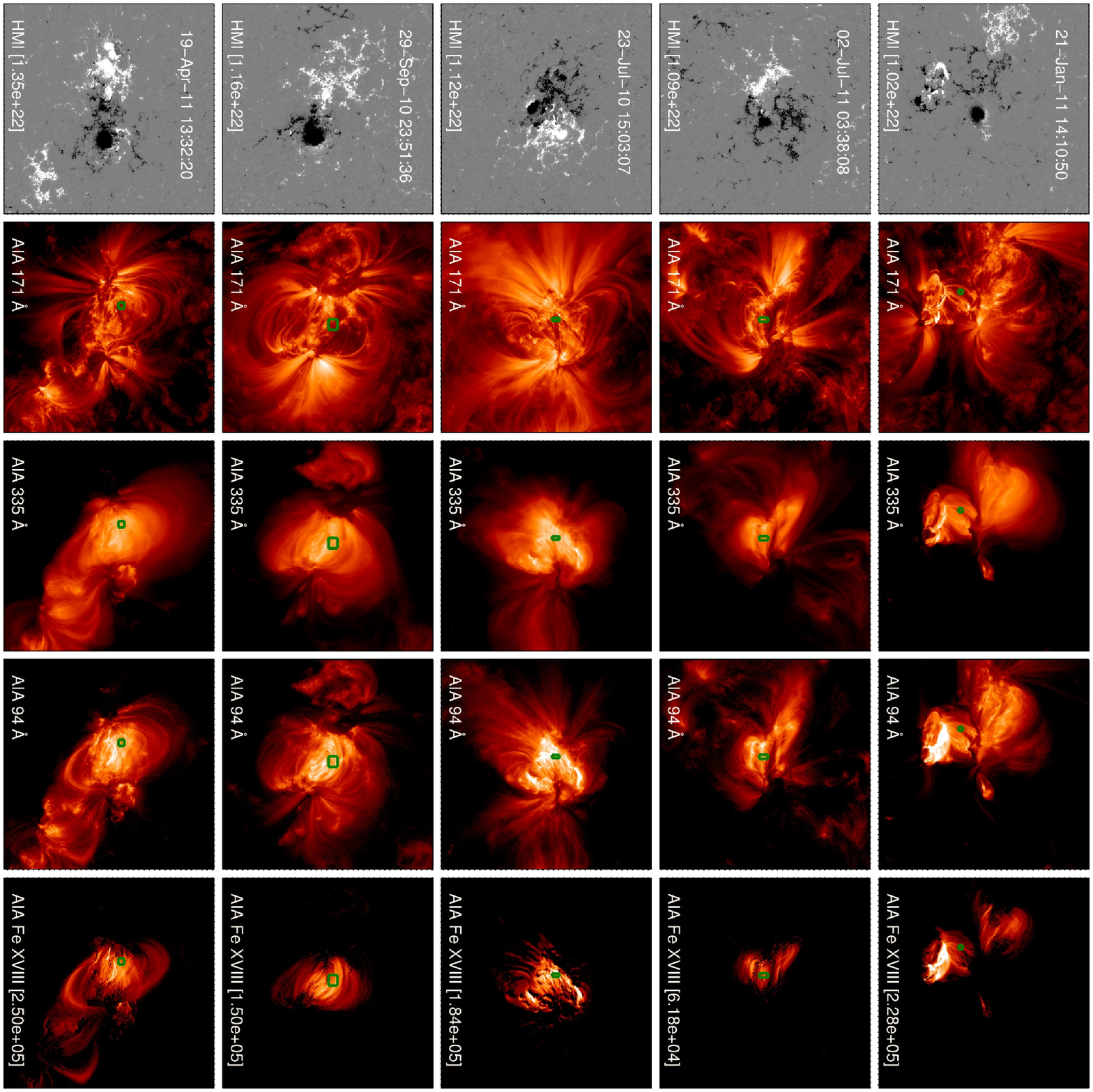}}
\caption{The same as Figure~\protect{\ref{fig:summary1}} but for regions 6--10 in
  Table~\protect{\ref{table:list}}.}
\label{fig:summary2}
\end{figure*}

\begin{figure*}[t!]
\centerline{\includegraphics[clip,angle=90,scale=0.67]{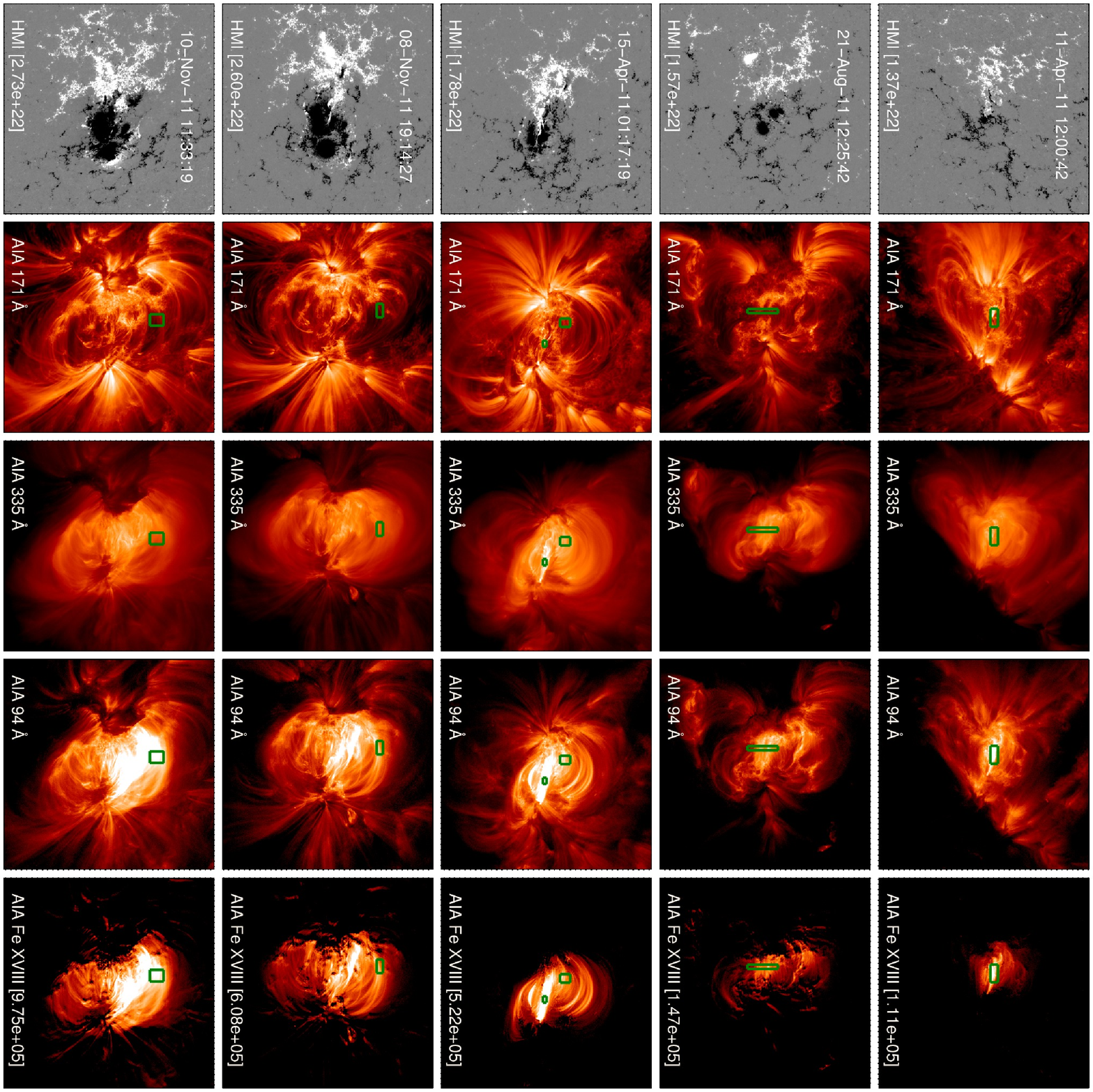}}
\caption{The same as Figure~\protect{\ref{fig:summary2}} but for regions 11--14 in
  Table~\protect{\ref{table:list}}.}
\label{fig:summary3}
\end{figure*}

In this paper we present a more systematic survey of active region core emission. It is
well known that the amount of high temperature emission scales with the total unsigned
magnetic flux \cite[e.g.,][]{schrijver1987} and we use this metric to parametrize the
observed active regions. We have selected 15 observations that span a wide range of
magnetic flux values ($10^{21}$--$10^{23}$\,Mx). For each region we compute the
differential emission measure (DEM) in the active region core using observations from the
EUV Imaging Spectrograph (EIS) and the Atmospheric Imaging Assembly (AIA). We focus on
intensities measured in the ``inter-moss`` region, that is, the region between the loop
footpoints where we are measuring the properties near the loop apex. Measurements of the
entire active region would potentially combine emission from both loop footpoints and
cooling loops and would require full models of the active region to interpret.

To better constrain the DEM at high temperatures we present a new method for isolating the
\ion{Fe}{18} emission in the AIA 94\,\AA\ channel. The results of this method have been
calibrated against spectroscopic observations of \ion{Fe}{18} 974.86\,\AA\
\citep{teriaca2012}. We find that for regions with appreciable magnetic flux the DEM in
the active region core is strongly peaked near 4\,MK consistent with our previous
results. For regions with weaker magnetic fields the amount of high temperature emission
diminishes significantly and the DEM becomes broader, consistent with the analysis of
\cite{tripathi2011} and \cite{viall2011}.

The observation of strongly peaked emission measure distributions in active region cores
casts strong doubts on the Parker nanoflare model of coronal heating \citep{parker1988} as
it is commonly interpreted.  The very small spatial scales expected for magnetic
reconnection relative to the 1\arcsec\ (725\,km) resolution of current coronal instruments
suggests that observed coronal loops should be composed of many unresolved threads that
are various stages of heating and cooling. This implies that the observed temperature
distributions should be broad \citep[e.g.,][]{cargill1994,klimchuk2001,cargill2004}. A
survey of hydrodynamic simulations of coronal loops by \cite{mulumoore2011} suggests that
for nanoflare heating models the temperature distribution has a power-law index of about 2
or less ($EM\sim T^\alpha$) Our analysis, in contrast, shows that the temperature
distribution in the core of an active region is often strongly peaked, with
$\alpha\sim$3--4. It is possible that magnetic reconnection in coronal loops behaves
differently than has been assumed. It could, for example, occur more frequently or on
larger spatial scales than previously imagined. At present, however, it appears that the
wave heating scenario suggested by \cite{vanballegooijen2011} is more easily reconciled
with the available observations.

\section{Observations}

\begin{deluxetable*}{rrrrrrrrrr}
\tabletypesize{\small}
\tablecaption{A Survey of Solar Active Regions\tablenotemark{a}}
\tablehead{
   \multicolumn{1}{c}{Region}      &
   \multicolumn{1}{c}{NOAA}        &
   \multicolumn{1}{c}{Date}        &
   \multicolumn{1}{c}{$X_{cen}$}   &
   \multicolumn{1}{c}{$Y_{cen}$}   &
   \multicolumn{1}{c}{$A_M$}       &
   \multicolumn{1}{c}{$\Phi_M$}    &
   \multicolumn{1}{c}{$I_{hot}$}   &
   \multicolumn{1}{c}{$\alpha$}    &
   \multicolumn{1}{c}{EIS File}
}
\startdata
     1 &     1082 &     19-Jun-10 01:57:44 & -306.4 &  439.3 & 2.87(19) & 4.08(21) & 1.36(04) &      2.2 & eis\_l1\_20100619\_014433 \\ 
     2 &     1158 &     12-Feb-11 15:32:13 & -248.4 & -211.8 & 3.04(19) & 4.22(21) & 1.88(04) &      2.7 & eis\_l1\_20110212\_143019 \\ 
     3 &     1082 &     21-Jun-10 01:46:37 &  162.9 &  405.2 & 3.29(19) & 4.72(21) & 1.05(04) &      2.0 & eis\_l1\_20100621\_011541 \\ 
     4 &     1259 &     25-Jul-11 09:36:09 &  224.7 &  323.4 & 3.98(19) & 5.68(21) & 7.81(03) &      2.0 & eis\_l1\_20110725\_090513 \\ 
     5 &     1150 &     31-Jan-11 11:25:19 & -470.9 & -250.6 & 5.17(19) & 7.50(21) & 1.47(05) &      2.2 & eis\_l1\_20110131\_102326 \\ 
     6 &     1147 &     21-Jan-11 14:10:50 &   26.6 &  476.5 & 6.49(19) & 1.02(22) & 2.28(05) &      3.6 & eis\_l1\_20110121\_133954 \\ 
     7 &     1243 &     02-Jul-11 03:38:08 & -299.0 &  216.6 & 6.79(19) & 1.09(22) & 6.18(04) &      2.9 & eis\_l1\_20110702\_030712 \\ 
     8 &     1089 &     23-Jul-10 15:03:07 & -363.4 & -453.6 & 6.96(19) & 1.12(22) & 1.84(05) &      3.5 & eis\_l1\_20100723\_143210 \\ 
     9 &     1109 &     29-Sep-10 23:51:36 &  361.5 &  261.5 & 7.19(19) & 1.16(22) & 1.50(05) &      4.3 & eis\_l1\_20100929\_223226 \\ 
    10 &     1193 &     19-Apr-11 13:32:20 &   36.3 &  363.5 & 7.94(19) & 1.35(22) & 2.50(05) &      3.3 & eis\_l1\_20110419\_123027 \\ 
    11 &     1190 &     11-Apr-11 12:00:42 & -492.6 &  281.0 & 8.61(19) & 1.37(22) & 1.11(05) &      3.0 & eis\_l1\_20110411\_105848 \\ 
    12 &     1271 &     21-Aug-11 12:25:42 &  -50.8 &  150.8 & 9.59(19) & 1.57(22) & 1.47(05) &      3.6 & eis\_l1\_20110821\_105251 \\ 
    13 &     1190 &     15-Apr-11 01:17:19 &  218.1 &  304.4 & 1.04(20) & 1.78(22) & 5.22(05) &  3.7,3.3 & eis\_l1\_20110415\_001526 \\ 
    14 &     1339 &     08-Nov-11 19:14:27 &   88.1 &  258.4 & 1.41(20) & 2.60(22) & 6.08(05) &      4.8 & eis\_l1\_20111108\_181234 \\ 
    15 &     1339 &     10-Nov-11 11:33:19 &  406.0 &  266.8 & 1.48(20) & 2.73(22) & 9.75(05) &      3.7 & eis\_l1\_20111110\_100028 
\enddata
\tablenotetext{a}{$X_{cen}$ and $Y_{cen}$ are the NOAA active region coordinates
  differentially rotated to the mid-point of the EIS raster. $A_{M}$ is the total area
  occupied by pixels between 50 and 500\,G in cm$^2$. $\Phi_M$ is the total unsigned
  magnetic flux in Mx. $I_{hot}$ is the total AIA \ion{Fe}{18} intensity in the field of
  view in DN~s$^{-1}$. The parameter $\alpha$ is the slope of the emission measure
  distribution between $\log T$ 6.0 and 6.6.}
\label{table:list}
\end{deluxetable*}

\begin{figure}[t!]
\centerline{\includegraphics[clip,scale=0.55]{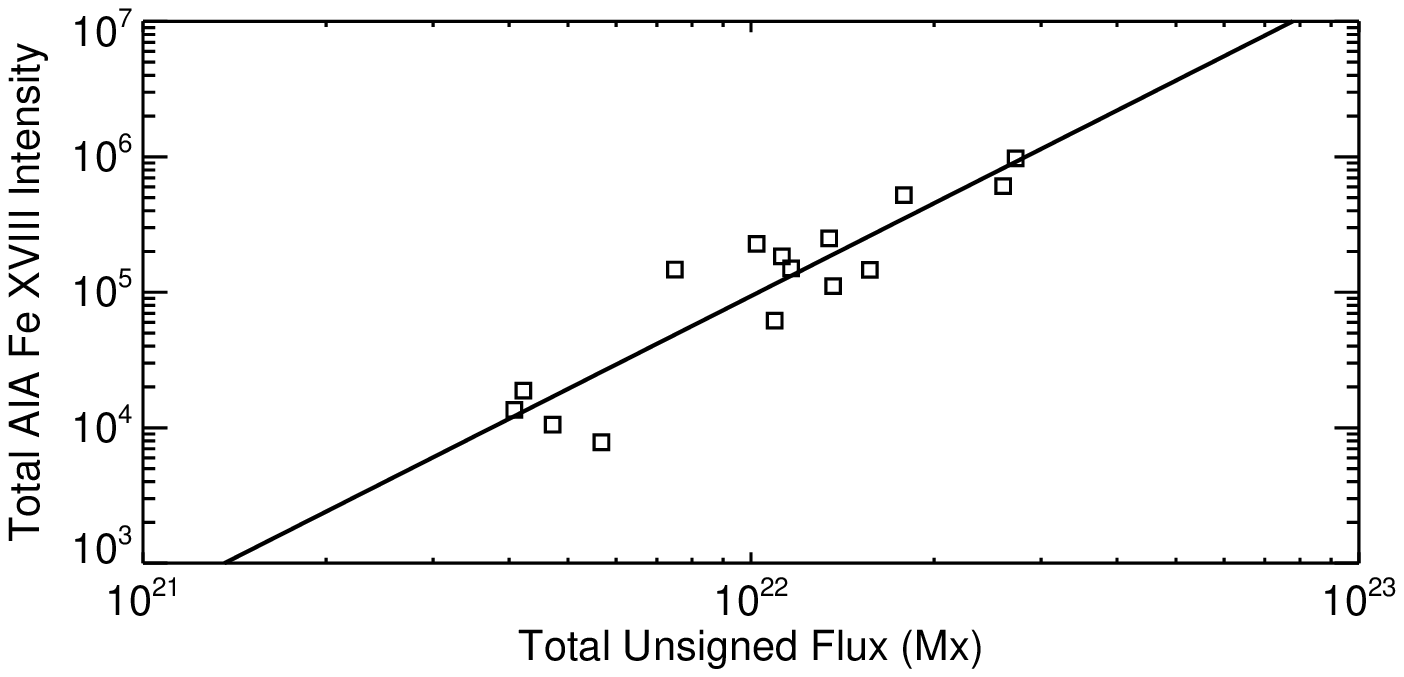}}
\caption{The total AIA \ion{Fe}{18} intensity ($I_{hot}$ in
  Table~\protect{\ref{table:list}}) as a function of the total unsigned magnetic flux
  ($\Phi_M$). The solid line is a power-law fit to the data. The total amount of high
  temperature emission in an active region varies strongly with the total unsigned
  magnetic flux.}
\label{fig:flux}
\end{figure}

Our aim here is to investigate the temperature structure of solar active regions
systematically. The observations of individual emission lines from the EIS instrument
provide detailed temperature diagnostics and introduces a strong constraint on our
analysis. EIS \citep{culhane2007,korendyke2006} is a high spatial and spectral resolution
imaging spectrograph. EIS observes two wavelength ranges, 171--212\,\AA\ and
245--291\,\AA, with a spectral resolution of about 22\,m\AA\ and a spatial resolution of
about 1\arcsec\ per pixel.  Solar images can be made by stepping the slit over a region of
the Sun and taking an exposure at each position.

Telemetry constraints generally limit the spatial and spectral coverage of an
observation. These constraints necessitate the selection of a limited number of spectral
windows in a raster and not all EIS observations include all of the potentially useful
emission lines. We have designed several EIS observing sequences that contain all of the
emission lines needed to compute emission measure distributions. Of particular importance
are the observation of emission lines from \ion{Ca}{14}--\ion{Ca}{17}, which constrain the
analysis at temperatures above 3\,MK \citep{warren2008b}. These studies have been run
frequently and we used summary images to manually review the available data and select a
set of observations that appeared to span a wide range of solar conditions. Each EIS
raster was processed in the usual way to remove the CCD pedestal and dark current,
identify any defective pixels, and calibrate the data. Intensities were then determined
for each emission line of interest at every spatial pixel using Gaussian fits to the line
profiles.

For each observation we determined the NOAA coordinates for the active region of interest
at the mid-point of the EIS raster. NOAA region numbers, times, and solar coordinates are
given in Table~\ref{table:list}. There are EIS observations taken during the interval
considered by \cite{viall2011}, but they do not include several of the high temperature Ca
lines and are not optimal for emission measure analysis. For completeness we have included
an EIS raster from this time. The observations analyzed by \cite{tripathi2011} pre-date
the launch of \textit{SDO} and are not included here. 

For each observation we obtained full-disk AIA images \citep{lemen2012} and full-disk
Helioseismic and Magnetic Imager (HMI, \citealt{scherrer2012}) magnetograms from the
Stanford JSOC data center. AIA is a set of multi-layer telescopes capable of imaging the
Sun at high spatial resolution (0.6\arcsec\ pixels) and high cadence (typically
12\,s). Images are available at 94, 131, 171, 193, 211, 304, and 335\,\AA. AIA images are
also available at UV and visible wavelengths, but they are not used in this analysis. HMI
also images the full Sun at high spatial resolution (0.5\arcsec\ pixels) and high cadence
(typically 45\,s). To simplify the data management we selected all of the data within
300\,s of the raster mid-point (the dates and times given in Table~\ref{table:list}),
processed the images with the standard \verb+aia_prep+ software to provide a common plate
scale, and averaged the images at each wavelength together. For the AIA EUV images we
divide each image by the exposure time. We then extracted a $400\arcsec\times400\arcsec$
region centered on the NOAA active region coordinates. For each magnetogram the line of
sight magnetic field is corrected to the radial value by dividing by the cosine of the
helospheric angle. Representative images are shown in
Figures~\ref{fig:summary1}--\ref{fig:summary3}. For each observation we coaligned the EIS
\ion{Fe}{12} 195.119\,\AA\ raster with the AIA 193\,\AA\ image.

For each HMI magnetogram we compute the total unsigned flux ($\Phi_M$) for radial magnetic
field strengths between 50 and 500\,G. The lower bound excludes the quiet Sun and the
upper bound excludes sunspots. These limits were used by \cite{warren2006} to study the
relationship between the total unsigned flux and the total soft X-ray intensity. As in
previous studies, they found a power-law relationship between the total intensity and the
magnetic flux, $I_{sxr}\sim\Phi_M^b$, with $b\approx1.6$. The values for the total
unsigned magnetic flux we find here are similar to those from our earlier study, which
used magnetogram data from the Michelson Doppler Imager (MDI) instrument on \textit{SoHO}
\citep{scherrer1995}. Note that the absolute magnetic fluxes measured with HMI need to be
reduced by a factor of 1.4 to agree with those measured with MDI \citep{liu2012}.

\begin{figure*}[t!]
\centerline{\includegraphics[clip,scale=0.5]{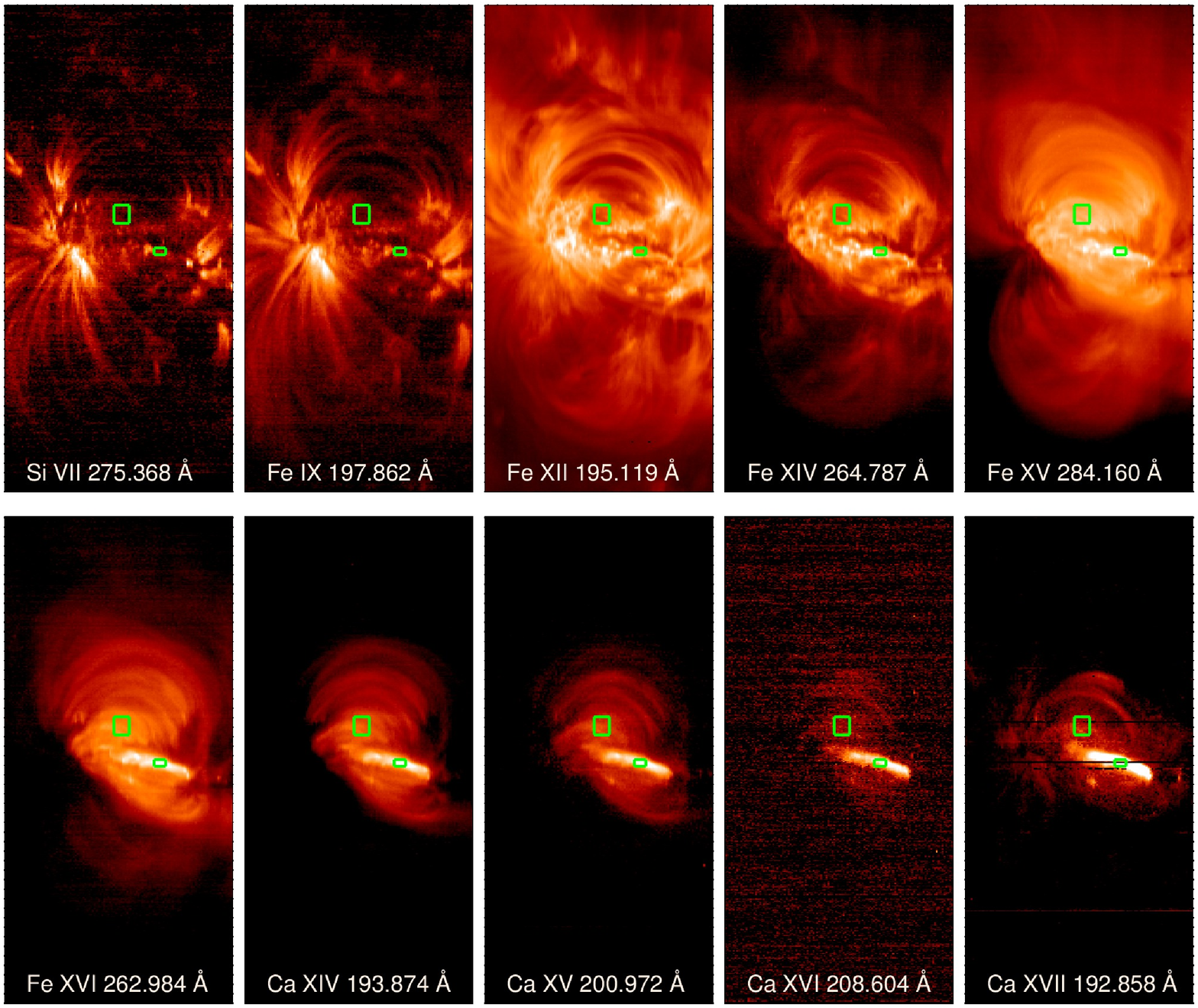}}
\hspace{-0.2in}\centerline{\includegraphics[clip,scale=0.6,angle=90]{f06.eps}}

\hspace{-0.2in}\centerline{\includegraphics[clip,scale=0.6,angle=90]{f07.eps}}
\caption{Example EIS active region observations. Selected rasters and line profiles are
  shown for the 15 April 2011 active region. The profiles from the intense ``bar'' of
  emission are displayed in the lower panels. The profiles from the larger box are
  displayed in the top panels. The red curves indicate the Ca lines of interest. The blue
  curves indicate emission lines formed at lower temperatures. Below each profile the
  difference between the fit and the observed line profile are displayed. The vertical
  line indicates the 1-$\sigma$ error for the intensity in the spectral bin.}
\label{fig:eis}
\end{figure*}

The 94\,\AA\ channel on AIA contains the \ion{Fe}{18} 93.92\,\AA\ line, which is one of
the most intense \ion{Fe}{18} transitions \citep[e.g.,][]{desai2005}. Since this line is
formed at a high temperature (7.1 MK) we expect that its integrated intensity will have a
dependence on $\Phi_M$ similar to that of the soft X-ray emission. Unfortunately, as can
be seen in Figures~\ref{fig:summary1}--\ref{fig:summary3} this wavelength range also
contains emission lines formed at lower temperatures. The atomic data for this wavelength
range is incomplete \citep{odwyer2010,testa2012}, further complicating the quantitative
use of this channel. As we describe in detail in Appendix~\ref{sec:appendixA}, it is
possible to use a combination of AIA 171 and 193\,\AA\ images to estimate the amount of
contaminating emission in the 94\,\AA\ channel empirically. Subtracting the estimated warm
emission from the observed 94\,\AA\ image isolates the \ion{Fe}{18} 93.92\,\AA\
contribution. We have applied our algorithm to each active region observation of interest
and the results are shown in the final columns of
Figures~\ref{fig:summary1}--\ref{fig:summary3}. Comparisons with spectroscopic
observations of \ion{Fe}{18} 974.86\,\AA\ are given in \cite{teriaca2012}.

For each AIA \ion{Fe}{18} image we have computed the total intensity in the active region
above 2\,DN~s$^{-1}$, which we consider to be the noise level introduced by the
subtraction method, and list these values in Table~\ref{table:list}. In
Figure~\ref{fig:flux} we show a plot of the total intensity as a function of total
unsigned magnetic flux. We also show a power law fit to the data and obtain a power-law
index of 2.3, somewhat higher than our previous result using soft X-ray images. This
exercise confirms that while our set of active region observations is not large, it does
sample a wide range of solar conditions. The range of total \ion{Fe}{18} intensity varies
by almost 2 orders of magnitude, from $10^4$ to $10^6$\,DN~s$^{-1}$.

The next step in this analysis is to manually select a small ``inter-moss'' region for
each active region. These sub-regions were chosen if they were bright in AIA \ion{Fe}{18}
but did not contain significant footpoint (moss) in AIA 171\,\AA. The term moss refers to
the footpoint emission of high temperature loops which appears bright in emission lines
formed near 1\,MK (see \citealt{berger1999} and references therein). We also attempted to
avoid 171\,\AA\ loop emission in the core of the active region, but for some observations
this was not possible. These selections are indicated by the boxes display in
Figures~\ref{fig:summary1}--\ref{fig:summary3}. Note that the inter-moss region considered
here for the 2010 July 23 active region is slightly different than that analyzed in
\cite{warren2011}. The highest AIA \ion{Fe}{18} intensities are seen in the 2011 April 15
active region as a bright ``bar'' of emission. For this active region we select two fields of
view, one on the bright bar and another where the intensities are weaker, but more similar
to the intensities observed in the other active regions.

\begin{figure*}[t!]
\centerline{\includegraphics[clip,scale=0.525]{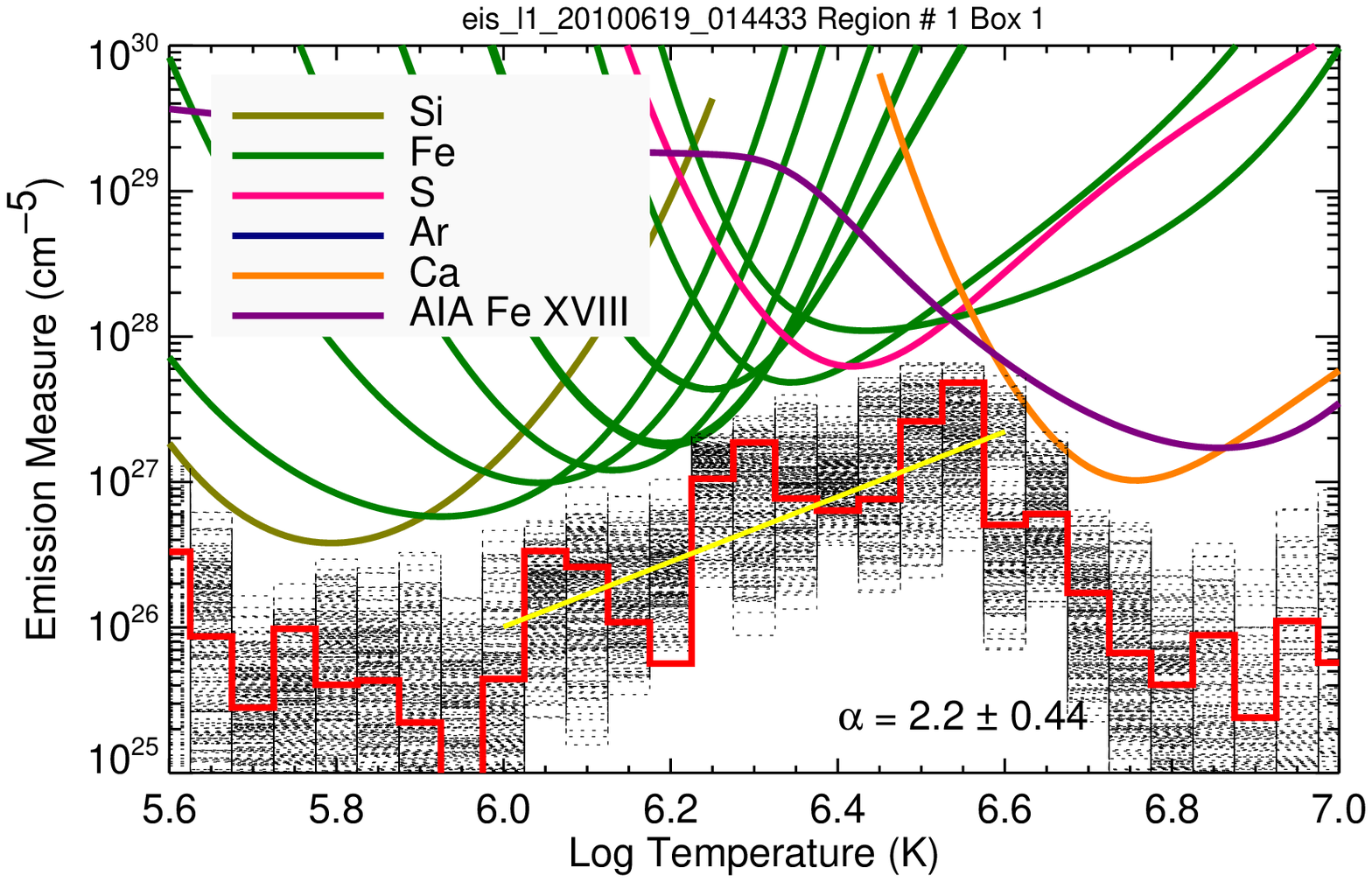}
            \includegraphics[clip,scale=0.525]{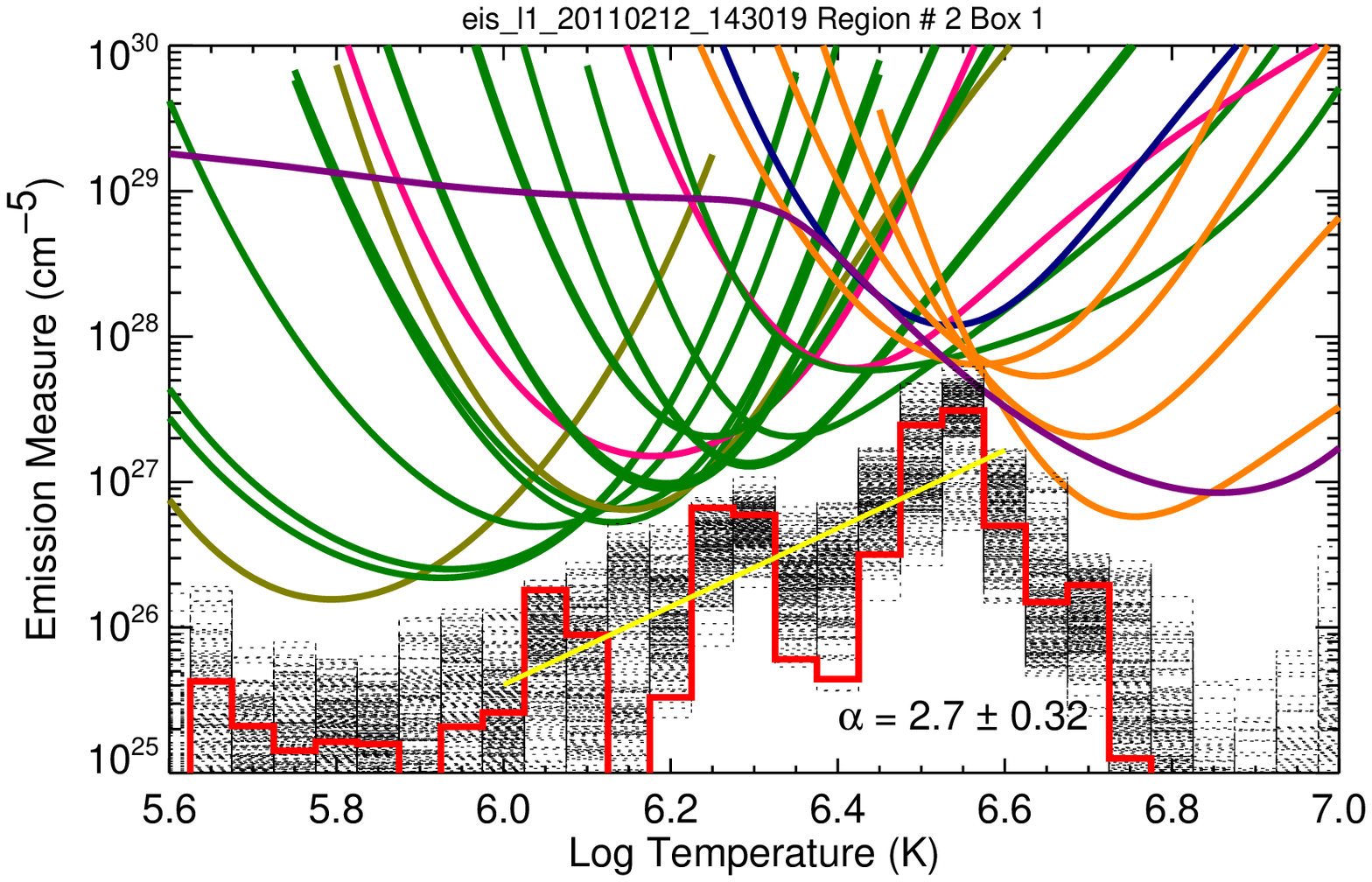}}
\centerline{\includegraphics[clip,scale=0.525]{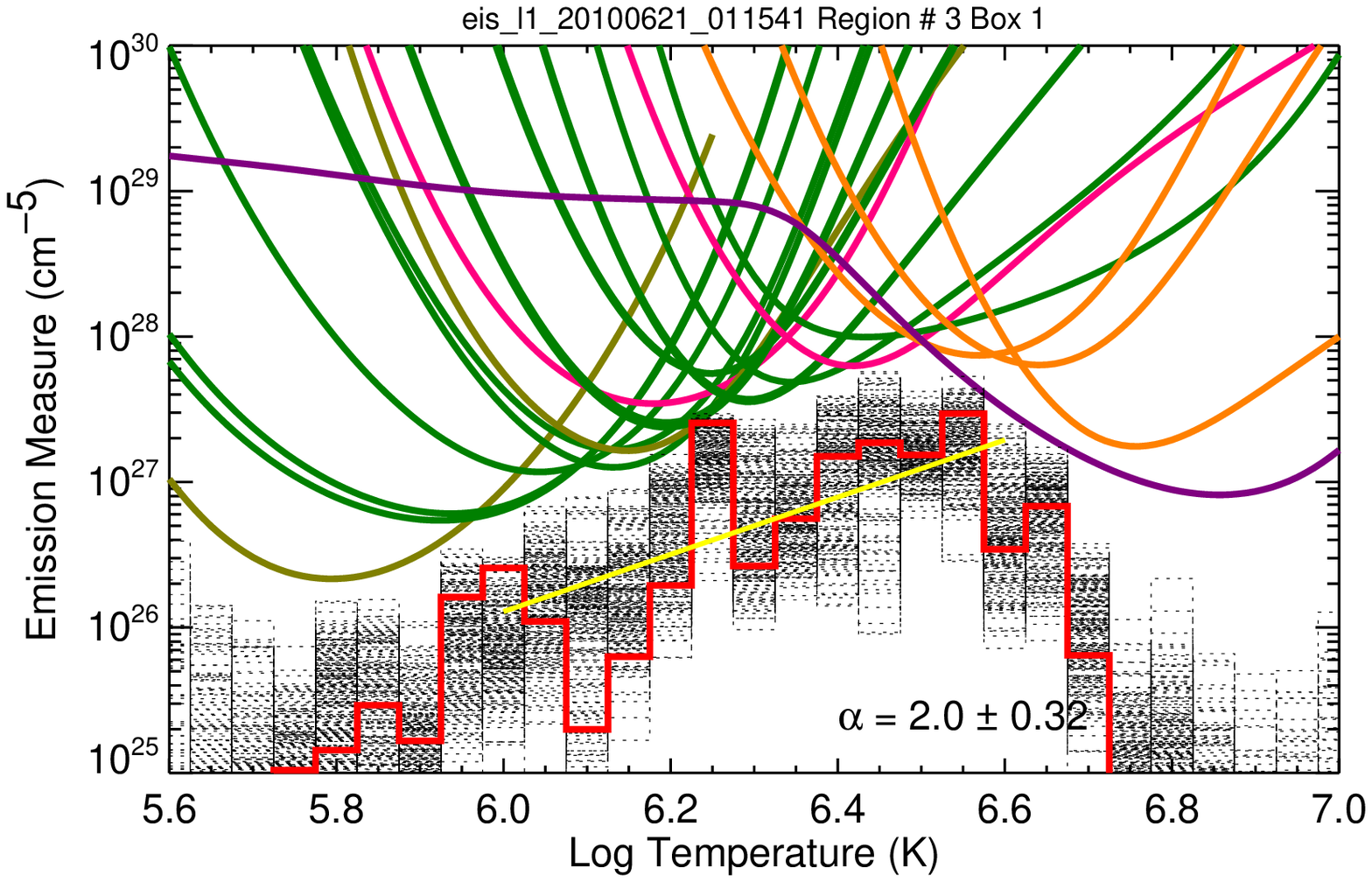}
            \includegraphics[clip,scale=0.525]{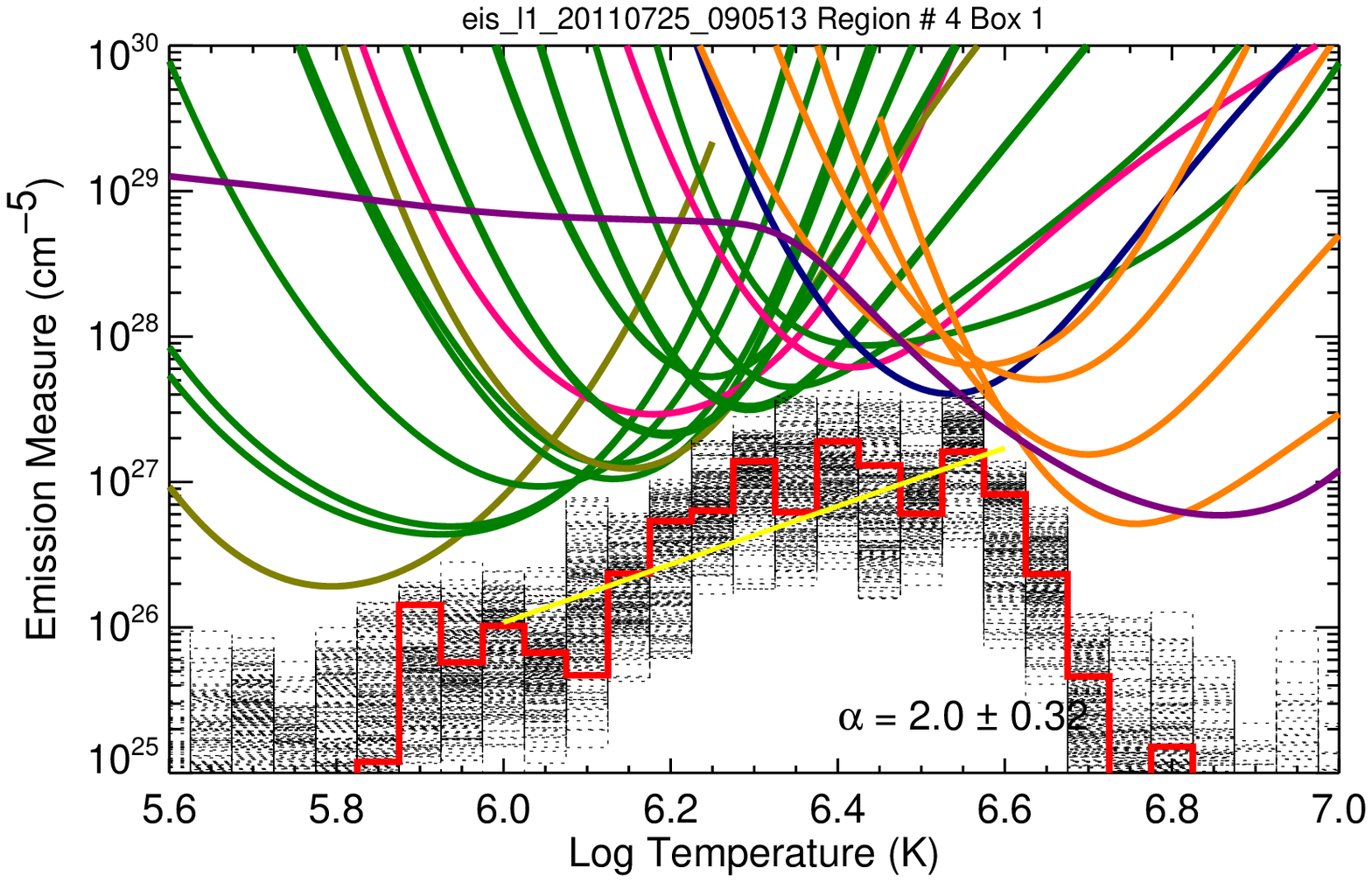}}
\centerline{\includegraphics[clip,scale=0.525]{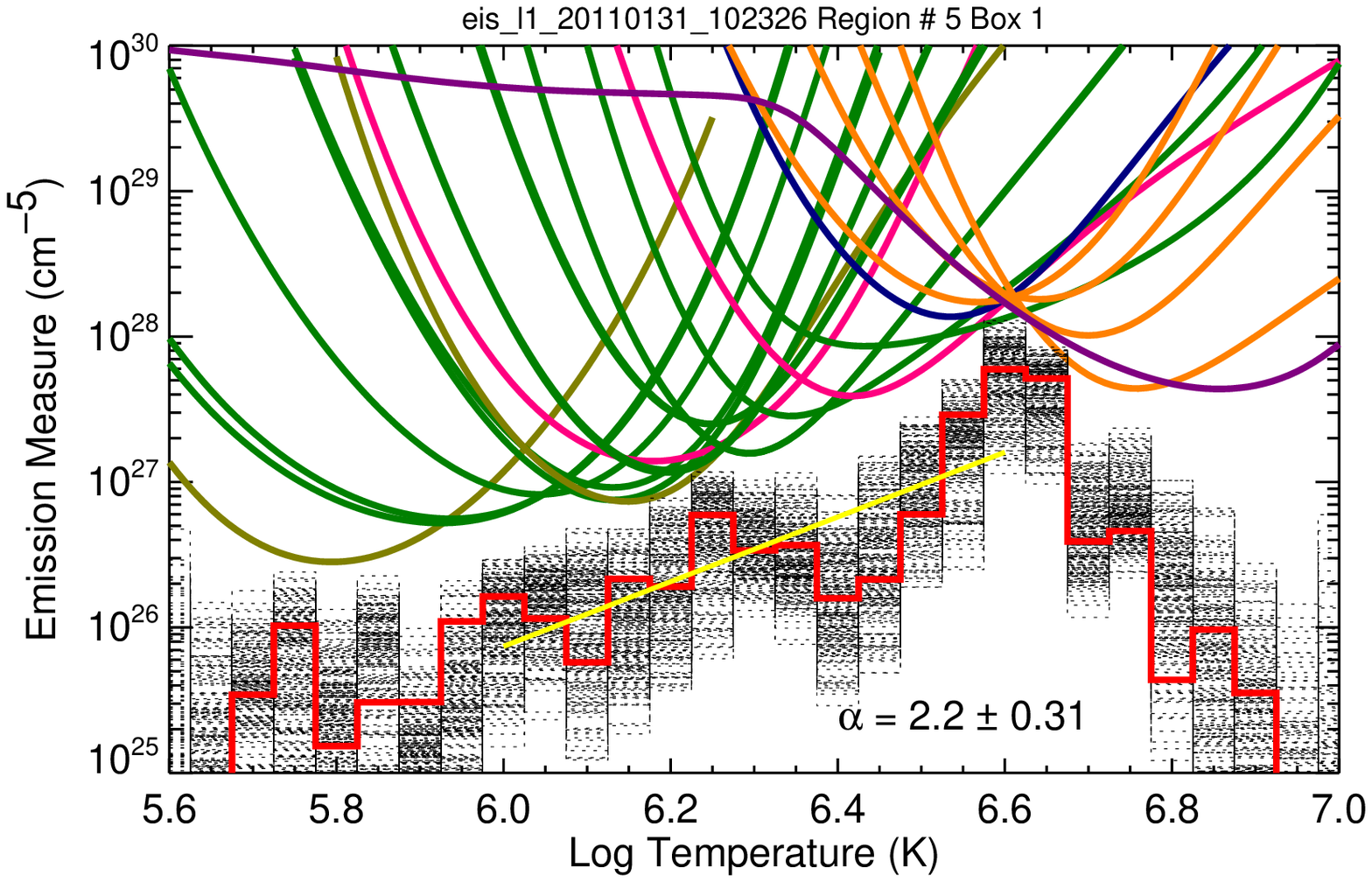}
            \includegraphics[clip,scale=0.525]{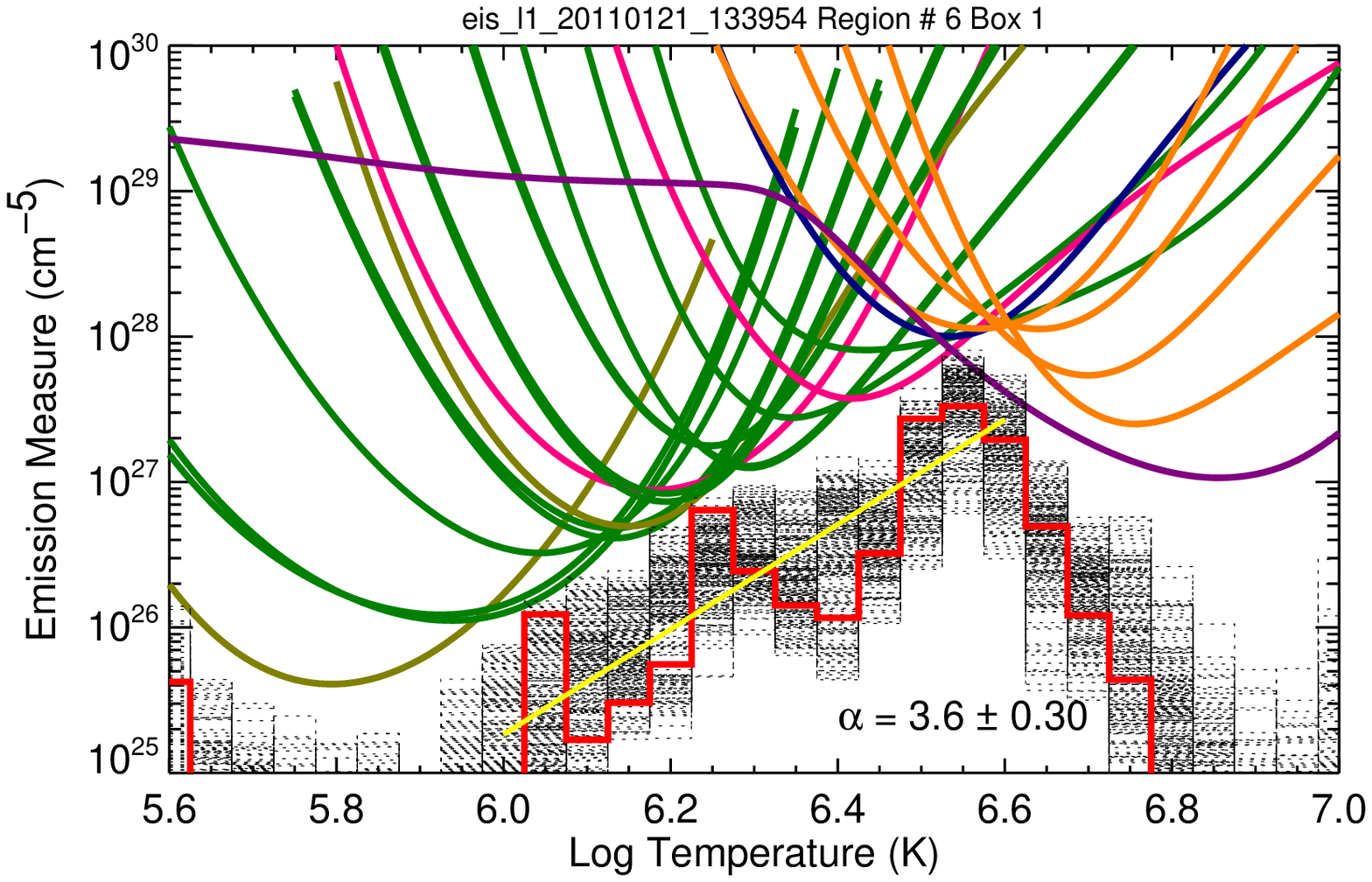}}
          \caption{The ``inter-moss'' emission measure distribution for 15 active
            regions. The field of view used to derived the intensities is indicated in
            each panel in
            Figures~\protect{\ref{fig:summary1}}--\protect{\ref{fig:summary3}}. For each
            measurement the emission measure distribution derived from MCMC is shown
            (thick red line) as well as the results from 250 Monte Carlo runs (black
            lines). The emission measure loci curves are color coded by ion. The slope of
            the distribution from 6.0 to 6.6 is indicated by the yellow line. The
            power-law index is also indicated on each plot. Regions 1--6 are shown here.}
\label{fig:dem1}
\end{figure*}

\begin{figure*}[t!]
\centerline{\includegraphics[clip,scale=0.525]{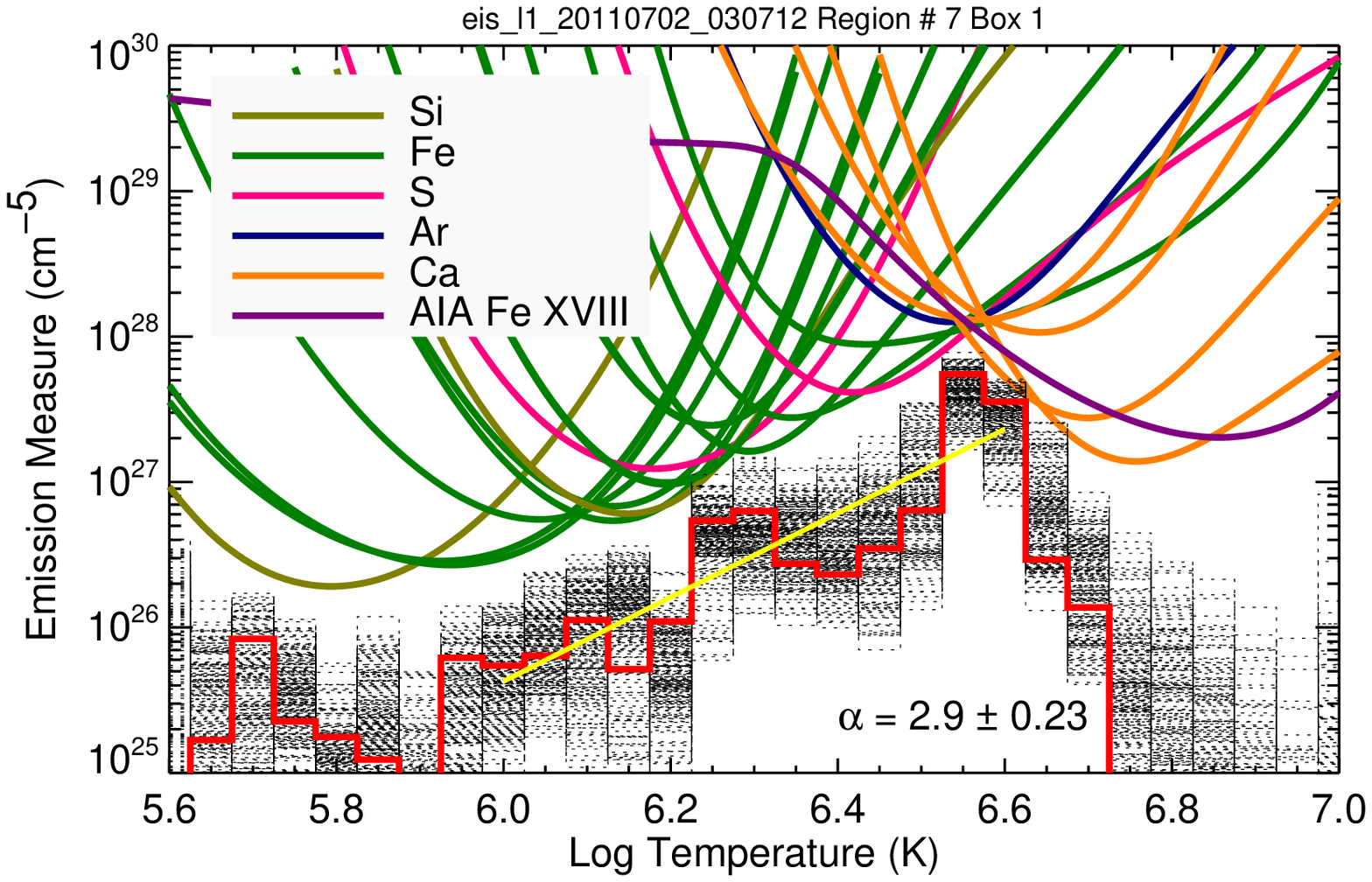}
            \includegraphics[clip,scale=0.525]{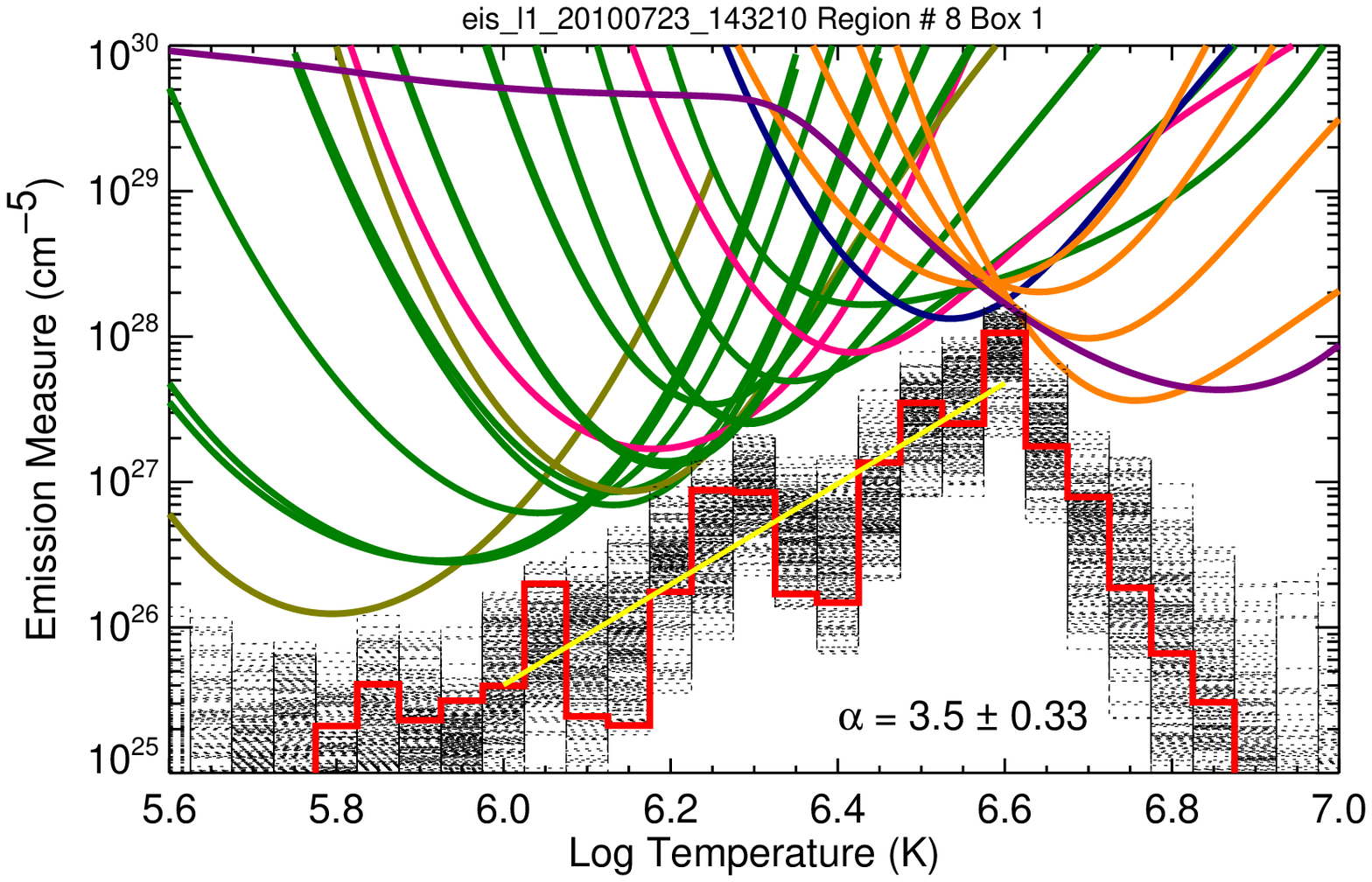}}
\centerline{\includegraphics[clip,scale=0.525]{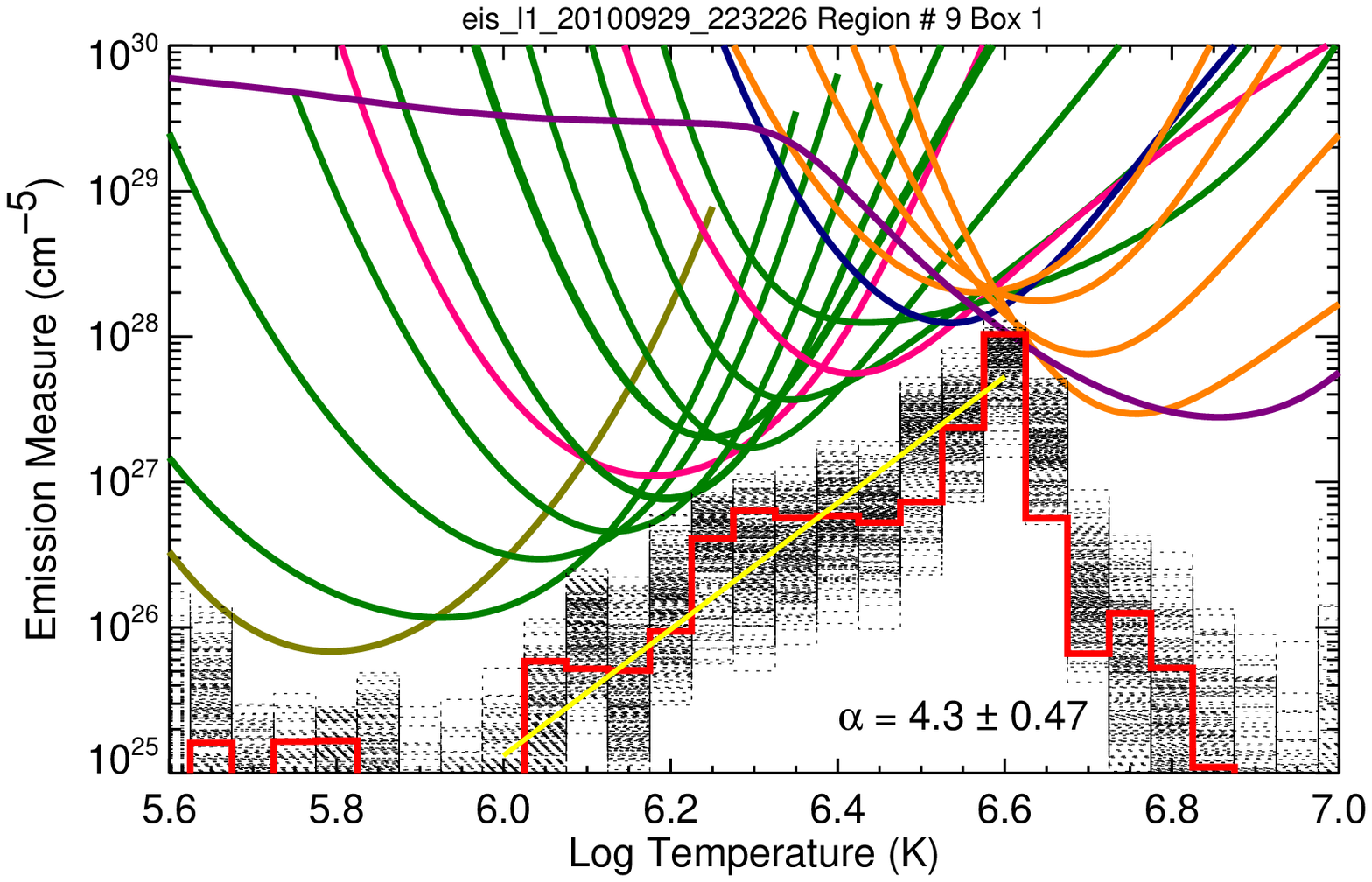}
            \includegraphics[clip,scale=0.525]{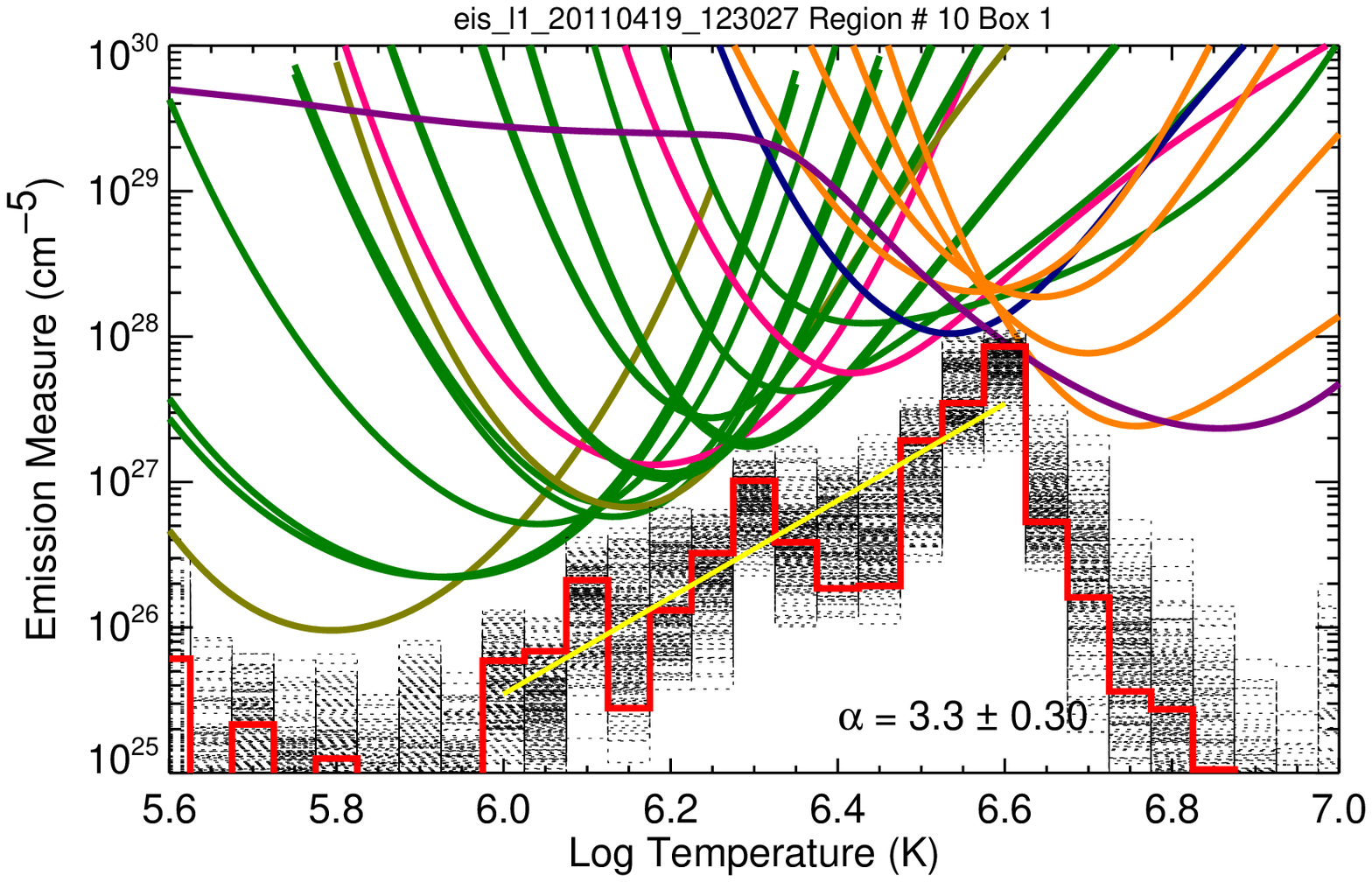}}
\centerline{\includegraphics[clip,scale=0.525]{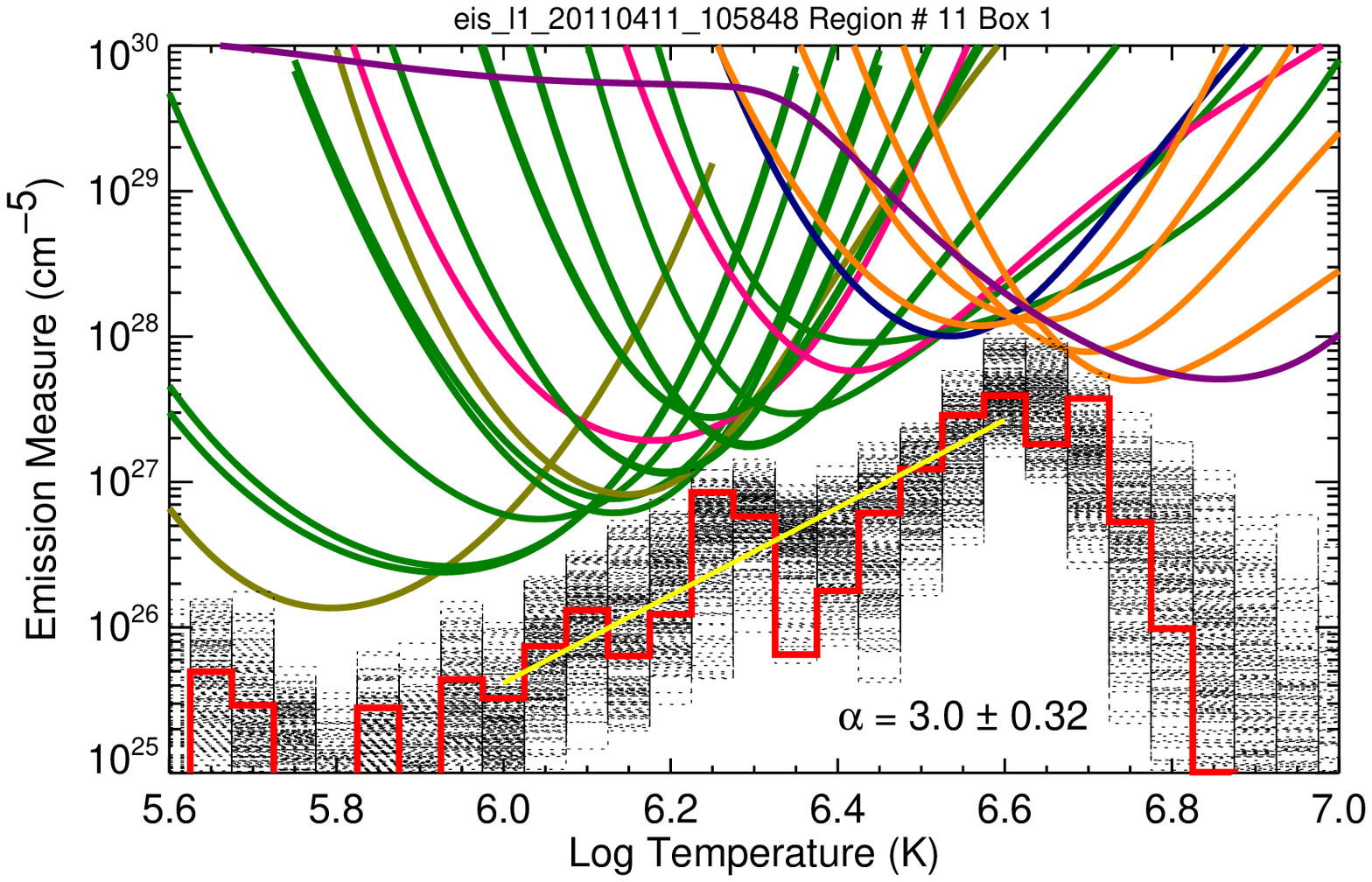}
            \includegraphics[clip,scale=0.525]{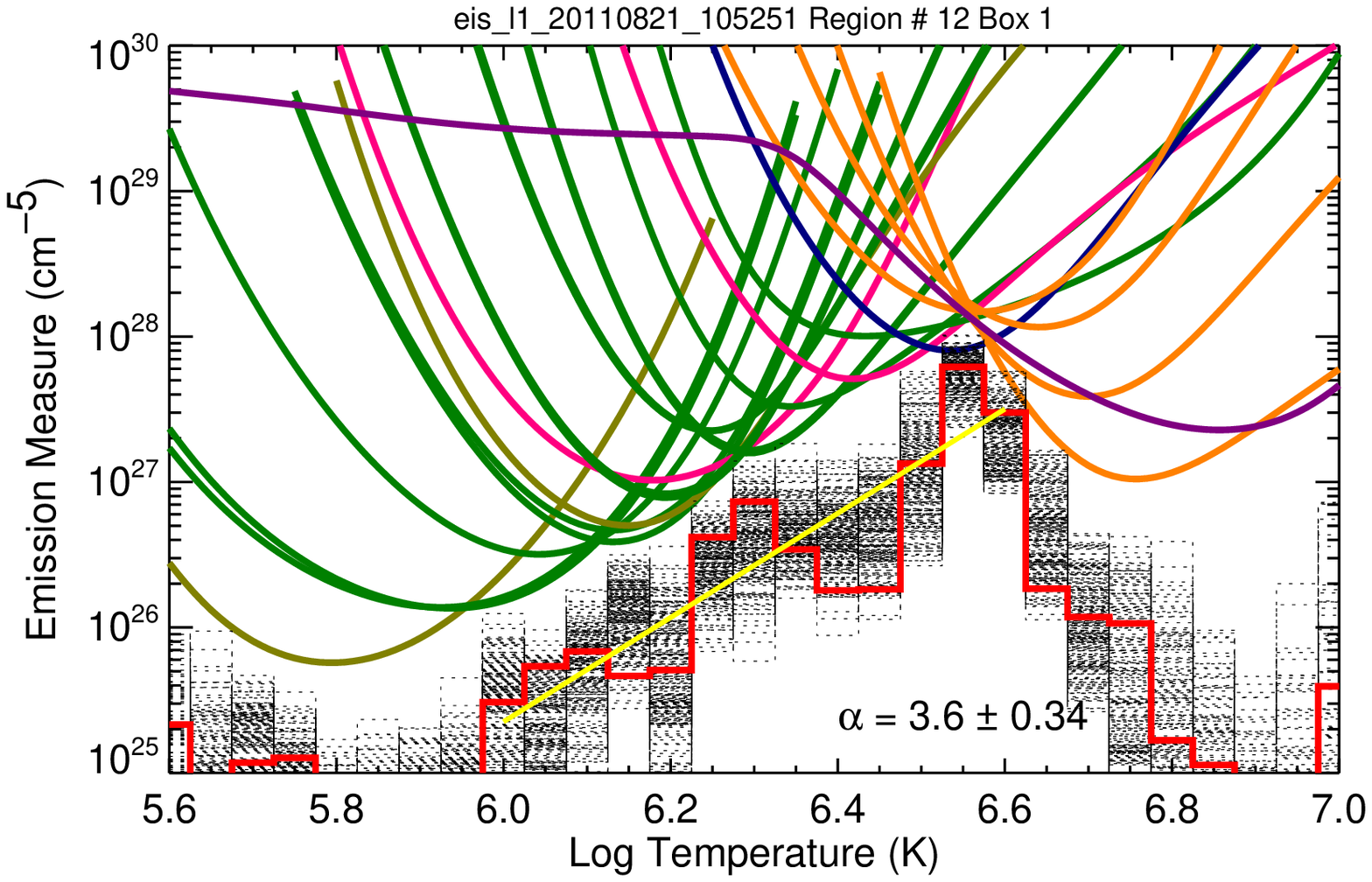}}
          \caption{The same as Figure~\protect{\ref{fig:dem1}} but for regions 7-12.}
\label{fig:dem2}
\end{figure*}

\begin{figure*}[t!]
\centerline{\includegraphics[clip,scale=0.525]{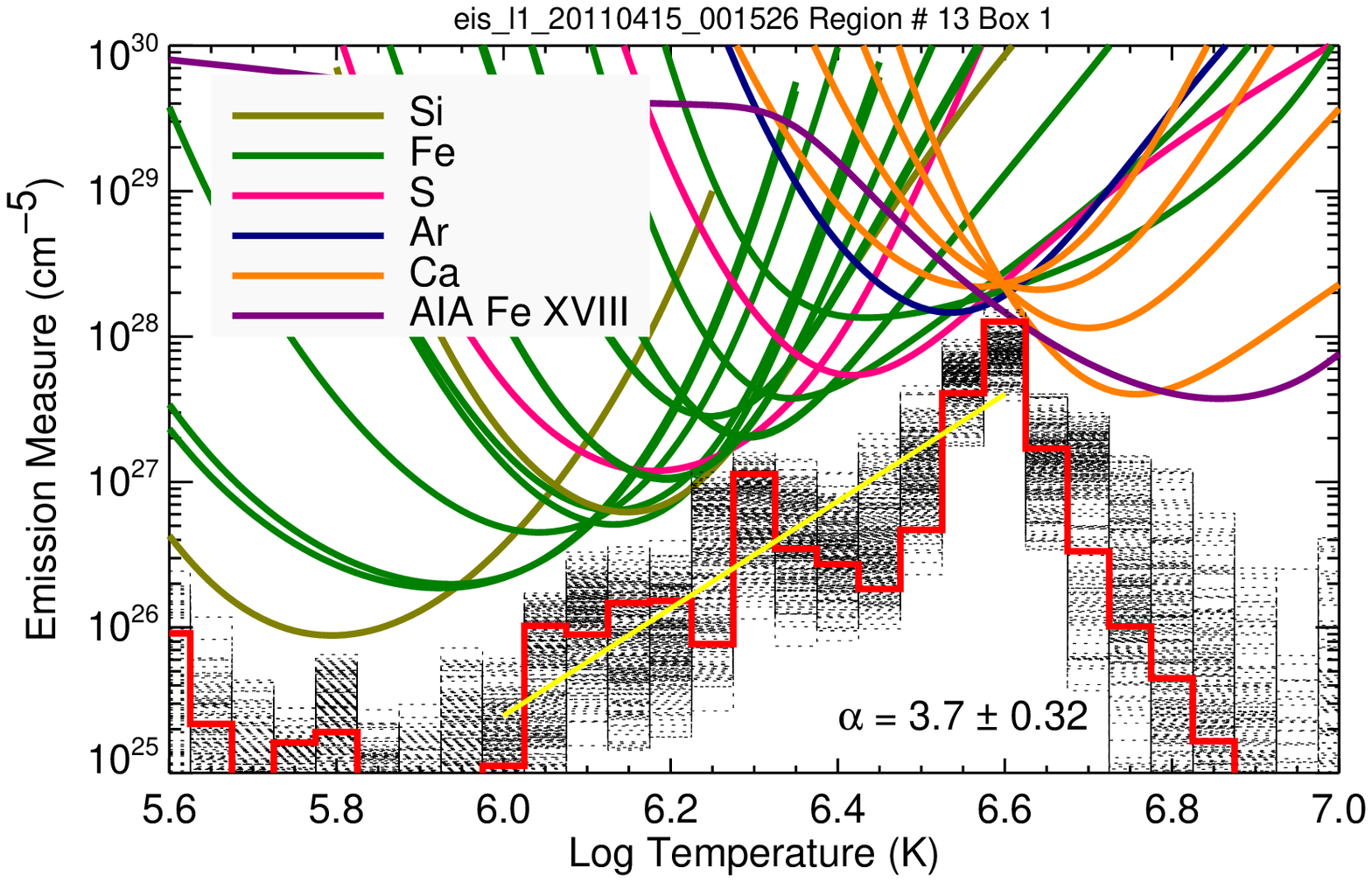}
            \includegraphics[clip,scale=0.525]{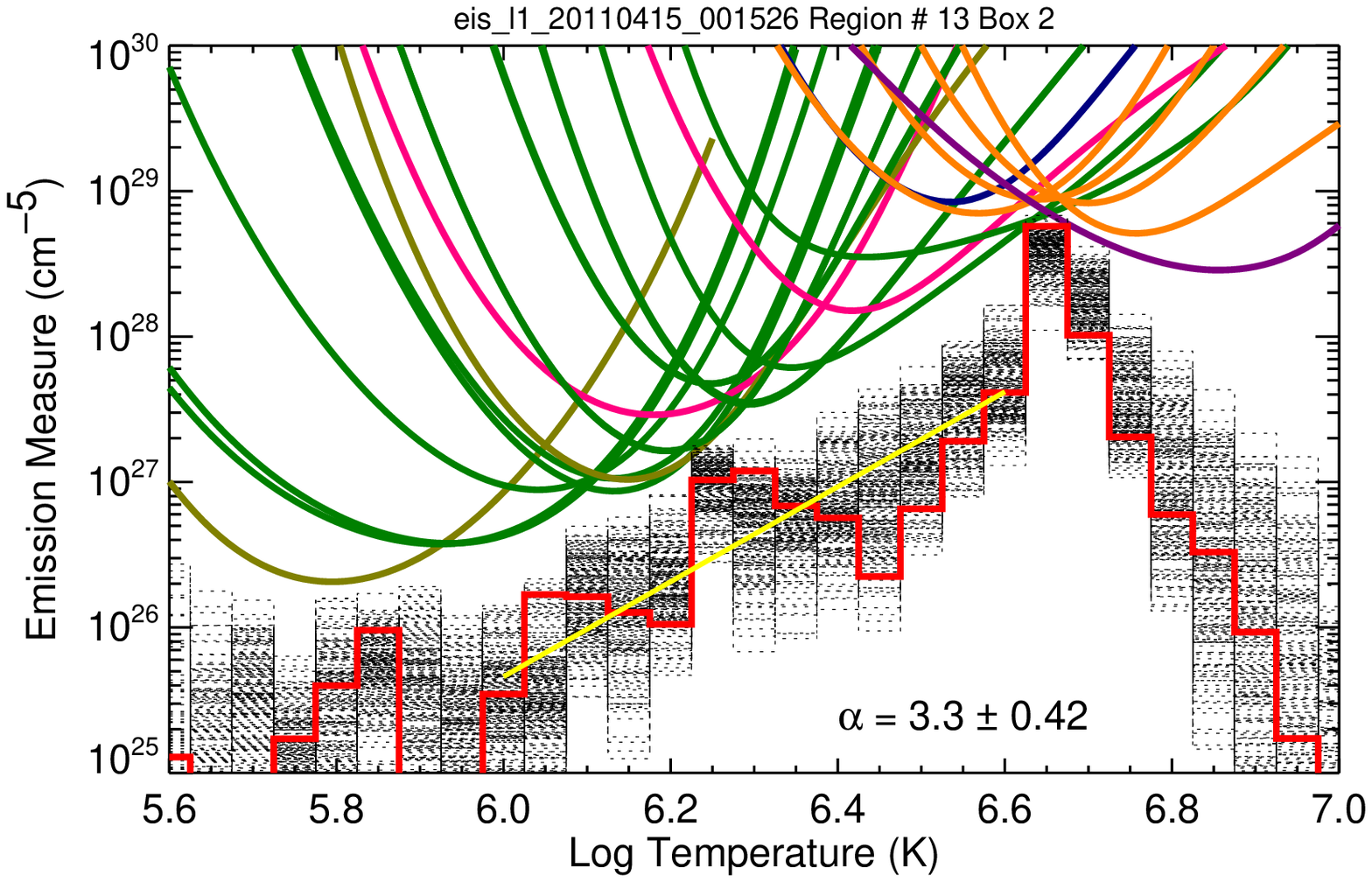}}
\centerline{\includegraphics[clip,scale=0.525]{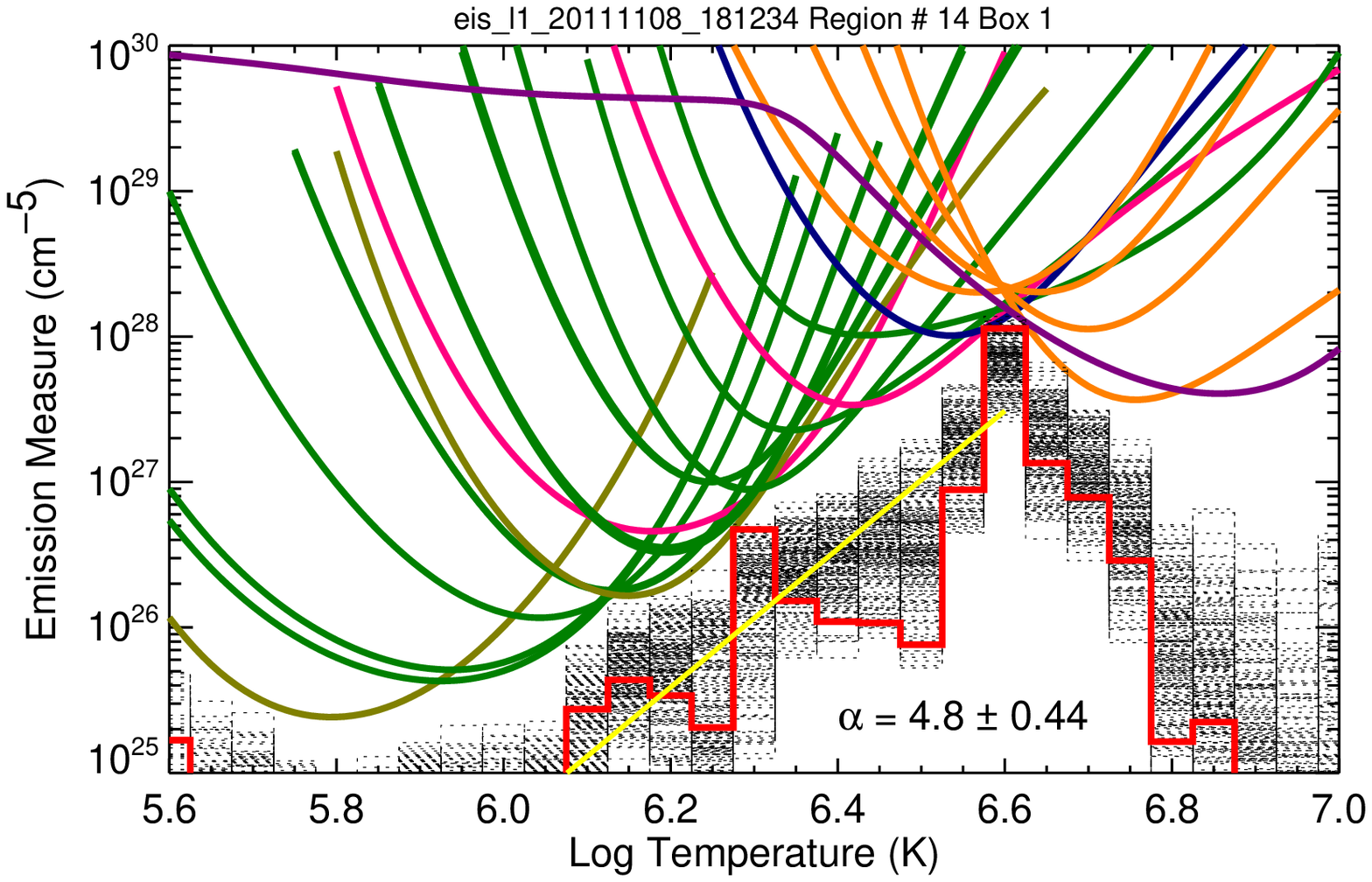}
            \includegraphics[clip,scale=0.525]{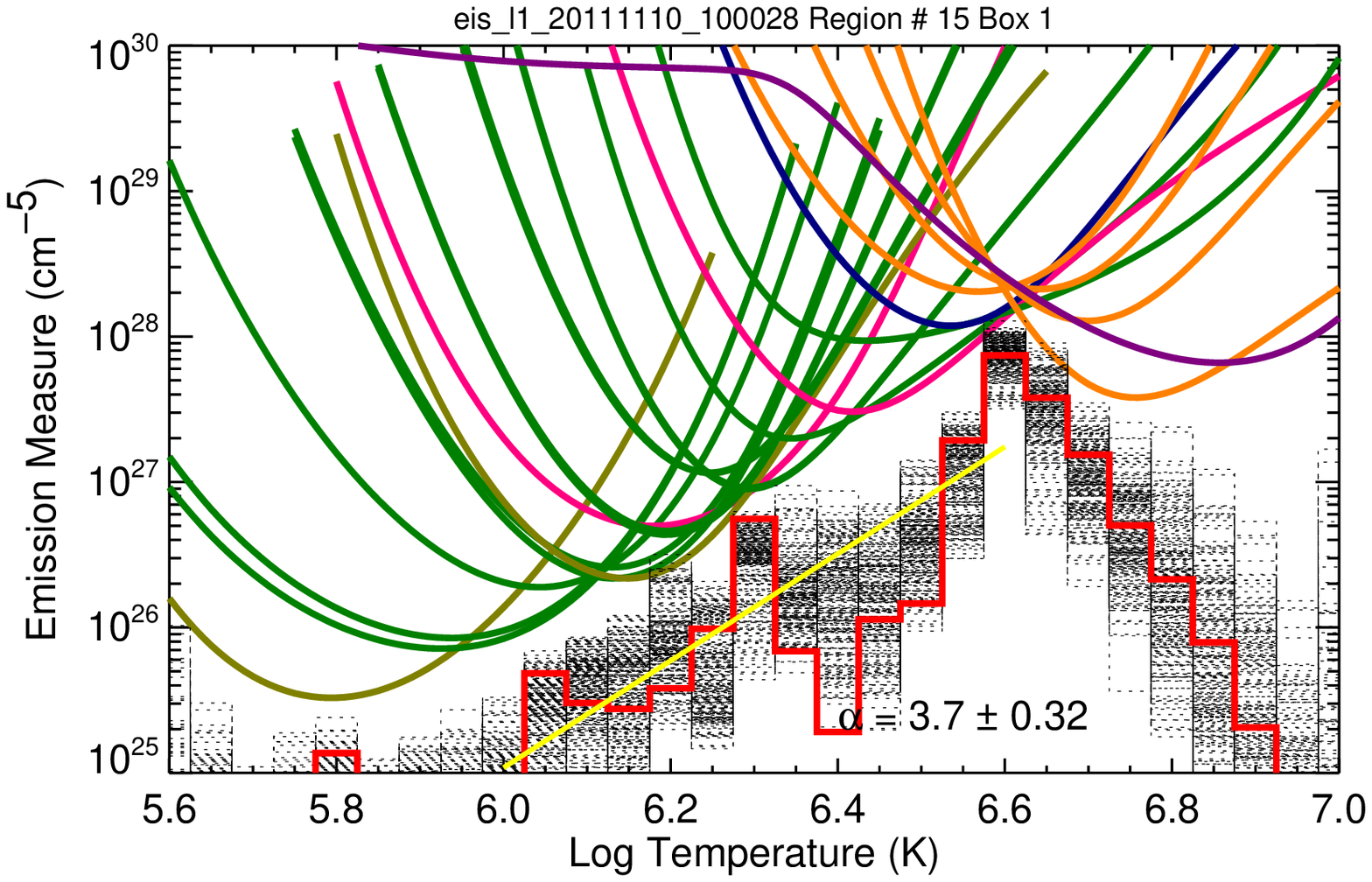}}
          \caption{The same as Figure~\protect{\ref{fig:dem1}} but for regions 13-15. Two
            emission measure distributions are shown for region 13.}
\label{fig:dem3}
\end{figure*}

For each inter-moss region we extracted all of the relevant EIS data from each spectral
window and averaged them together to create high signal-to-noise line profiles. In
computing these averaged profiles missing data are not included. We then fit the line
profiles with single Gaussians. The \ion{Ca}{17} 192.858\,\AA\ line is blended with
\ion{Fe}{11} 192.813\,\AA\ and a complex of \ion{O}{5} lines. We use the method outlined
by \cite{ko2009} to disentangle this blend. To insure consistency between the fits to the
Ca lines we use the widths measured for the \ion{Ca}{14} 193.874\,\AA\ and \ion{Ca}{15}
200.972\,\AA\ lines to constrain the fits to the other Ca lines. The width of the
\ion{Ca}{16} 2008.604\,\AA\ is set equal to that of \ion{Ca}{15} 200.972\,\AA. The width
of \ion{Ca}{17} 192.858\,\AA\ is limited to be within 0.05\,m\AA\ of the width of
\ion{Ca}{14} 193.874\,\AA. Example rasters, line profiles, and fits are shown in
Figure~\ref{fig:eis}.

The final EIS line list for each active region is generally the same used in
\cite{warren2011}, except that we now add intensities for \ion{Ar}{14} 194.396\,\AA, a
high first ionization potential (FIP) element that is useful for measuring the
composition. As before, we also include \ion{S}{10} 264.233\,\AA\ and \ion{S}{13}
256.686\,\AA, but Ar is formed at a somewhat higher temperature and has a higher FIP.
\cite{delzanna2012} have considered the relative intensities of some of these high FIP
lines in a diffuse off-limb active region spectrum and suggested potential problems with
blends. Our intensities for these lines are approximately 50 larger and we are able to
obtain consistent results for these lines. The \ion{Ca}{14}--\ion{Ca}{16} lines are not
available for the 2010 June 19 active region, which is included here because it was
studied by \cite{viall2011}.

\begin{figure}[t!]
\centerline{\includegraphics[clip,scale=0.55]{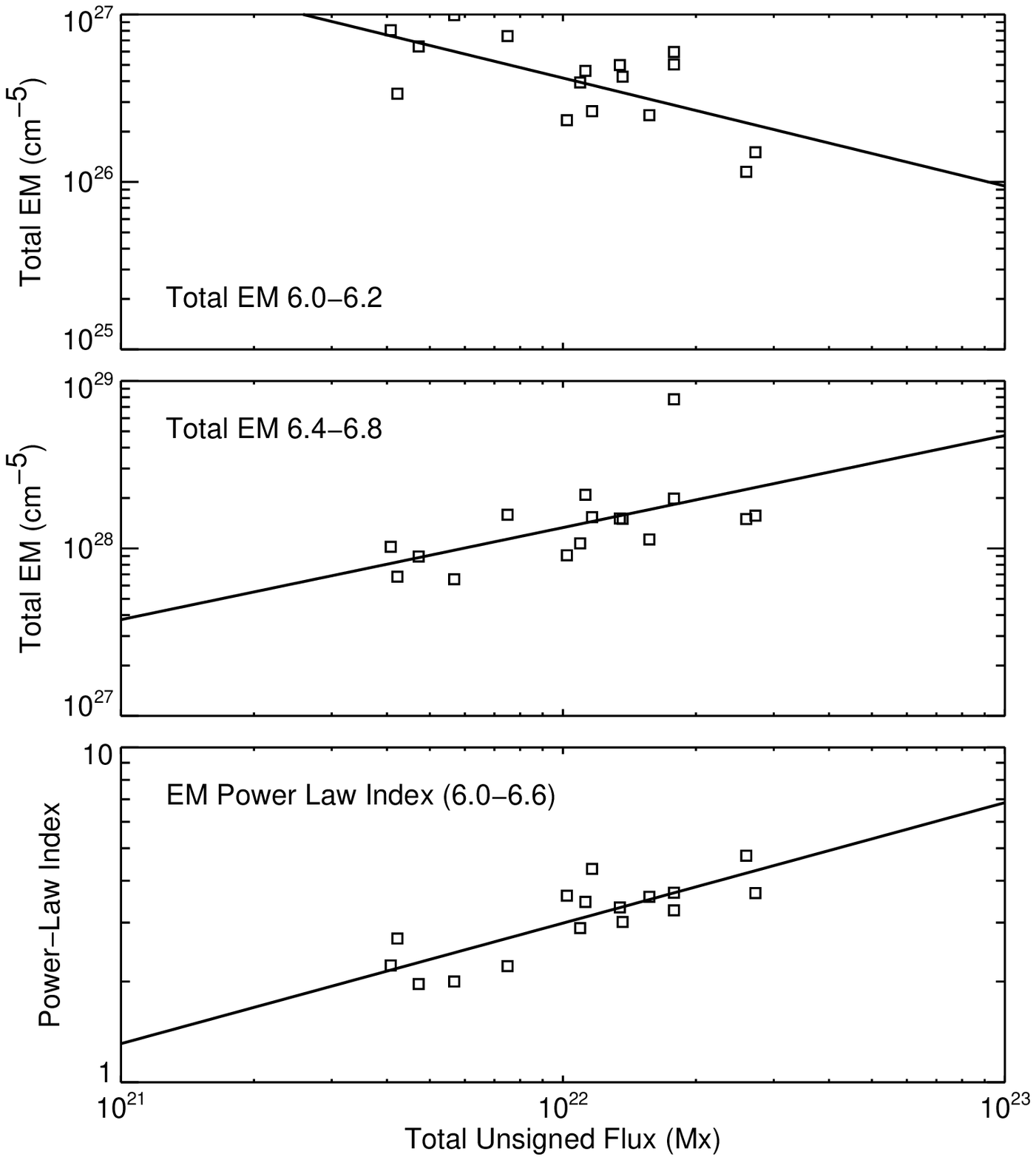}}
\caption{The total inter-moss emission measure at ``warm'' temperatures ($\log T_e=$
  6.0--6.2) and hot temperatures ($\log T_e=$ 6.4--6.8) as a function of the total
  unsigned magnetic flux. The bottom panel shows the power law index ($\alpha$) on the EM
  distribution between $\log T_e$ of 6.0 and 6.6.}
\label{fig:flux2}
\end{figure}

The intensity that we observe with EIS is related to line emissivity and the emission
measure distribution by the usual expression
\begin{equation}
   I_\lambda = \frac{1}{4\pi}\int \epsilon_\lambda(n_e,T_e)\xi(T_e)\,dT_e,
   \label{eq:ints}
\end{equation}
where $\epsilon_\lambda(n_e,T_e)$ is the emissivity computed with the CHIANTI atomic
database version 7 assuming coronal abundances \citep{feldman1992} and the CHIANTI
ionization fractions \citep{dere2009}. The function $\xi(T_e)=n_e^2\,ds/dT_e$ is the
differential emission measure distribution and the challenge we face is to infer this
distribution from the observed intensities.  It is also useful to consider the emission
measure loci computed from
\begin{equation}
  \xi_{loci}(T_e) = \frac{4\pi I_\lambda}{\epsilon_\lambda(n_e,T_e)},
\end{equation}
which indicates the temperature range where the various lines are sensitive. Note that to
aid in the comparisons with the em loci we will always plot the DEM multiplied by the
temperature bin,
\begin{equation}
  \xi(T_e)\,dT_e,
\end{equation}
and we refer to this as the emission measure distribution (EM). 

We also wish to use AIA \ion{Fe}{18} intensities derived from our subtraction method to
further constrain the emission measure calculations. The hottest strong emission line
observed with EIS during non-flaring conditions is \ion{Ca}{17} 192.858\,\AA, which is
formed at about 5\,MK. As mentioned previously, this line is blended, which adds
considerable uncertainty to the intensity. To utilize the subtracted AIA \ion{Fe}{18}
intensities we have computed a new response for this channel that only contains
contributions from \ion{Fe}{18} and continuum. The response distributed with the official
software contains contributions from several of the known emission lines formed at lower
temperatures.

To compute the differential emission measure we use the Monte Carlo Markov Chain (MCMC)
emission measure algorithm \citep{kashyap1998,kashyap2000} distributed with the
\verb+PINTofALE+ spectral analysis package. This algorithm has the advantage of not
assuming a functional form for the differential emission measure. The MCMC algorithm also
provides for estimates of the error in the EM by calculating the emission measure using
perturbed values for the intensities. The algorithm assumes the uncertainties in the
intensities are uncorrelated so that systematic errors in the calibration, which could
depend on the wavelength, or in the atomic data, which could vary by ion, are not
accounted for.

For each ``inter-moss'' field of view we have run the MCMC algorithm to compute the
differential emission measure. Additionally 250 Monte Carlo runs have also been performed
for each field of view. The resulting temperature distributions are shown in
Figures~\ref{fig:dem1}--\ref{fig:dem3}. The agreement between the observed intensities and
those computed from the EM is generally good, with most differences at the $\pm25\%$
level. For some of the weakest active regions the intensities of the hot Ca lines become
difficult to determine and the differences between the observed and computed intensities
are as much as 50\%. Inspection of quiet regions suggests that the Ca lines are all weakly
blended. Quiet sun intensities are typically about 5\% of the intensity in the core of a
very bright active region. For the weaker regions the impacts of the blends are more
significant. We have not attempted to correct for these blends and so the observed
intensities represent upper bounds.

Inspection of the emission measure distributions indicates that many are strongly peaked
near 4\,MK ($\log T_e = 6.6$), similar to the result from our previous analysis
\citep{winebarger2011,warren2011}. To quantify the steepness of the emission measure we
fit a power law of the form $EM\sim T^\alpha$ to each distribution. We have used two
methods to perform the fits. First we taken the median value of the emission measure in
each temperature bin from each Monte Carlo simulation and fit the resulting
distribution. We have also fit each distribution individually. Both methods yield
consistent results. The values for $\alpha$ are indicated on each plot as well as in
Table~\ref{table:list}. The uncertainties indicated in the plots are the 1-$\sigma$
standard deviations in the indexes determined from fitting each distribution and suggest
uncertainties of 10 to 20\%. In this sample 11 of the 16 EMs have $\alpha\ge3$. However,
we also measure 5 temperature distributions that are much shallower, with $\alpha\sim2$,
which is similar to the results from \cite{tripathi2011}. It is clear that these shallow
EMs are much more common in the active regions with the weakest magnetic fields.

Inspection of the emission measure distributions reveals an unexpected trend in the amount
of 1\,MK emission in the core of an active region. In regions 1 through 5 the emission
measure near 1\,MK is often between $10^{26}$ and $10^{27}$\,cm$^{-5}$. In the regions
with the strongest magnetic fluxes (regions 10--15) the emission measure appears to be
somewhat smaller, typically between $10^{25}$ and $10^{26}$\,cm$^{-5}$. To quantify this
we sum the emission measure between $\log T_e$ of 6.0 and 6.2 and plot it as a function of
total unsigned magnetic flux. As is indicated in Figure~\ref{fig:flux2} the emission
measure at these lower temperatures is inversely proportional to the field strength. This
clearly evident in Figures~\ref{fig:summary1}--\ref{fig:summary3}, which show relatively
few loops in the inter-moss regions in the 171\,\AA\ for the largest values of magnetic
flux.

The emission measure at the highest temperatures, as expected, rises with increasing total
unsigned magnetic flux. This is also shown in Figure~\ref{fig:flux2}. It is important to
recognize that this comparison between the properties of the inter-moss DEM and the total
unsigned magnetic flux is not ideal since we are comparing an apex property of selected
loops with the magnetic properties of the entire active region. As pointed out by
\cite{schrijver1987}, much of the increase in the total unsigned magnetic flux simply
reflects an increase in the area of the active region. The mean field strength also rises
with increasing active region area, but weakly (also see \citealt{fludra2008}). Ideally we
would compare the properties of the DEM with the magnetic properties at the loop
footpoints, but this would depend on having accurate methods for extrapolating the
photospheric field into the corona and such extrapolations have proven difficult to
achieve \citep{derosa2009}. It seems likely that trends observed Figure~\ref{fig:flux2}
would also be evident in a plot of EM as a function of footpoint field strength, but this
has yet to be demonstrated. 

\section{Discussion}

We have presented the calculation of emission measure distributions for 15 active region
observations spanning almost an order of magnitude in total unsigned magnetic flux. This
analysis suggests that the shape of the emission measure distribution depends on the
magnetic properties of the active region. For regions with appreciable magnetic flux the
emission measure distribution is often strongly peaked at a temperature of about
4\,MK. For lower levels of magnetic flux, however, we do observe shallower temperature
distributions. This suggests a possible resolution of the varied results presented
previously \citep{tripathi2011,viall2011,winebarger2011,warren2011}.

These results are difficult to reconcile with the Parker nanoflare model
\citep{parker1988}, at least as it has often been interpreted
\citep{cargill1994,klimchuk2001,cargill2004}. As mentioned previously, hydrodynamic
simulations suggest much flatter emission measure distributions than we observe in most of
these active regions \citep{mulumoore2011}. In the simulations the steepest slopes
($2.0\le\alpha\le2.3$) are obtained for radiative losses based on coronal abundances. For
all of the inter-moss regions that we considered the intensities of the S and Ar emission
lines computed from the DEM are consistent with what is observed, indicating that our
assumption of coronal abundances is correct. It is possible, however, that some of the
assumptions made in the hydrodynamic simulations, such as constant loop cross section or
the highly simplified chromosphere, produce misleading results.

It seems likely that high frequency heating that is concentrated at low heights in the
solar atmosphere will be able to account for the active region properties that we present
here.  The wave heating model described in \cite{vanballegooijen2011} and
\cite{asgari-targhi2012} appears to be a viable candidate. Detailed simulations, however,
are required to establish this.

\begin{figure}
\centerline{\includegraphics[clip,scale=0.525]{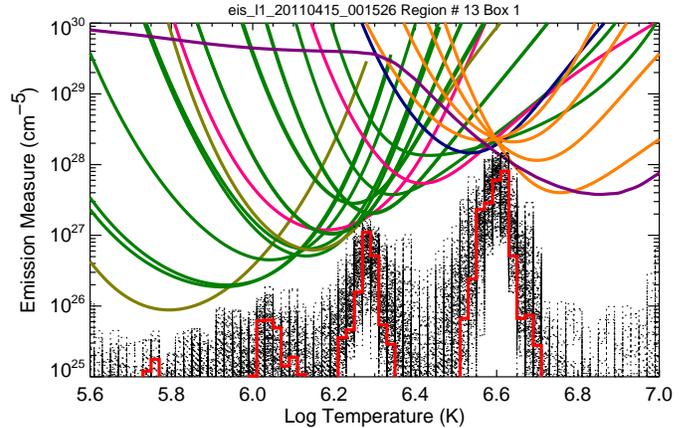}}
\caption{An alternative emission measure distribution derived for region 1 of the 2011
  April 15. This DEM was computed using a much smaller bin size ($\Delta\log T_e = 0.02$)
  than those shown in Figures~\ref{fig:dem1}--\ref{fig:dem2} ($\Delta\log T_e = 0.05$).}
\label{fig:dem_alt}
\end{figure}

\begin{figure*}[t!]
\centerline{\includegraphics[clip,angle=90,scale=0.67]{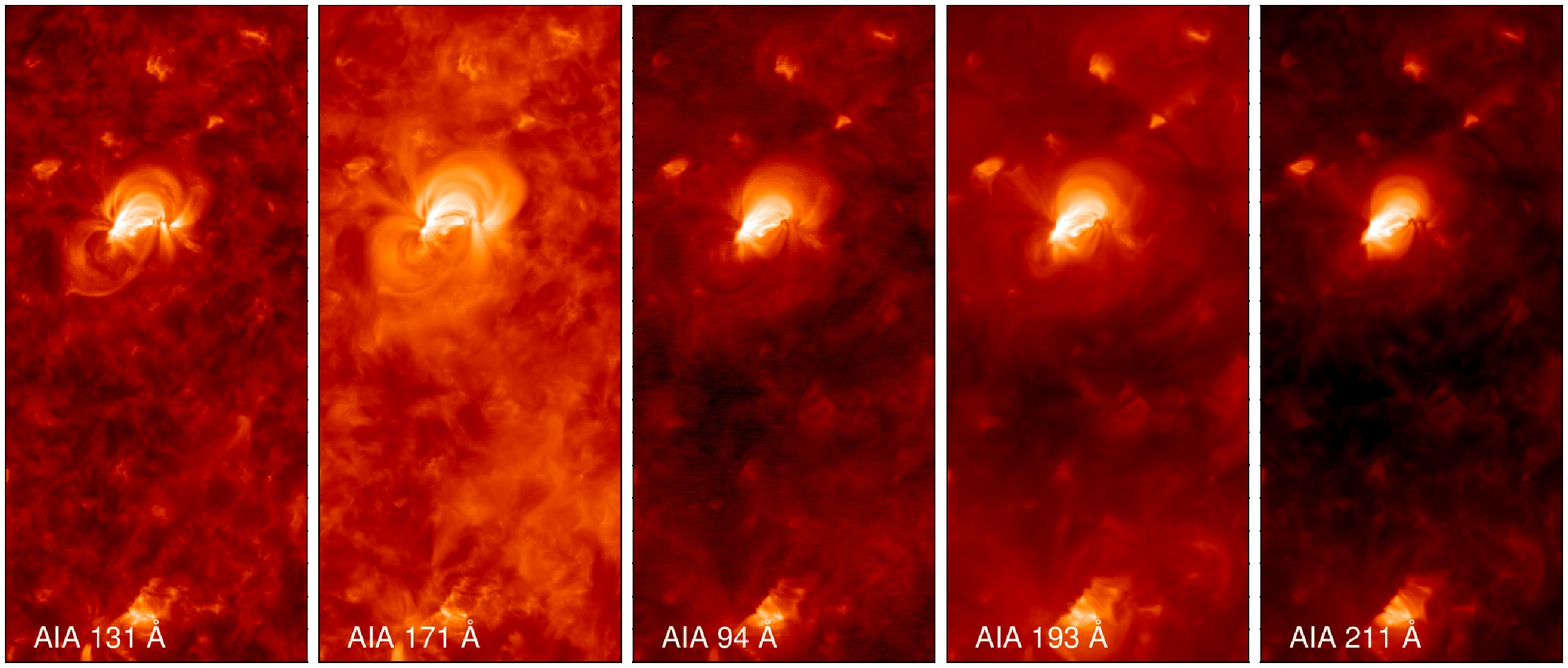}}
\centerline{\includegraphics[clip,angle=90,scale=0.67]{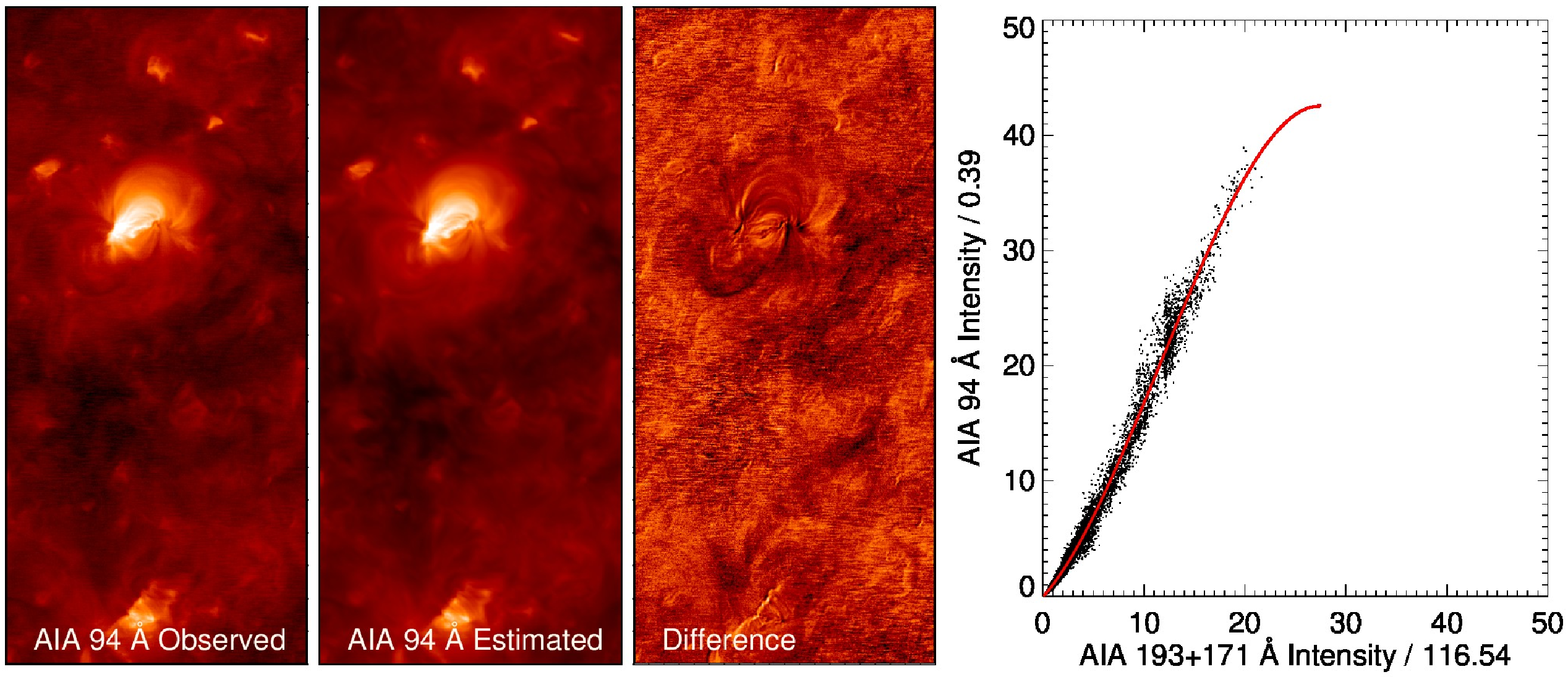}}
\caption{Time-averaged AIA images of a bright point and the quiet Sun. Each image is an
  average of 300 co-aligned exposures. These data were taken 2010 March 22 from 12 to 13
  UT. The bottom panels show the observed AIA 94\,\AA\ image compared with the intensities
  inferred from 171 and 193\,\AA. The differences between these images is also shown. The
  far right panel shows a polynomial fit to the intensities.}
\label{fig:qs1}
\end{figure*}

\begin{figure*}[t!]
\centerline{\includegraphics[clip,angle=90,scale=0.67]{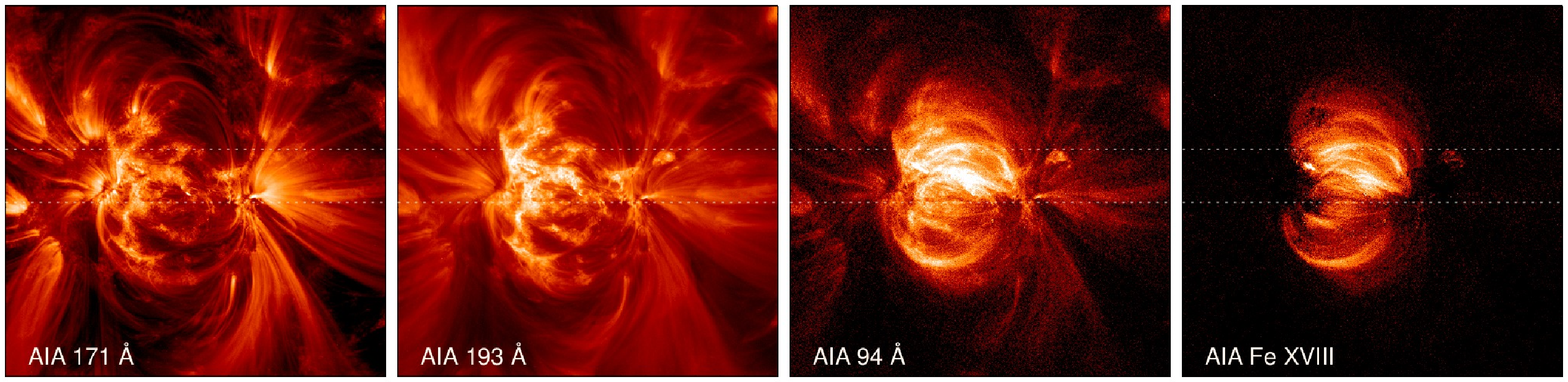}}
\centerline{\includegraphics[clip,angle=90,scale=0.67]{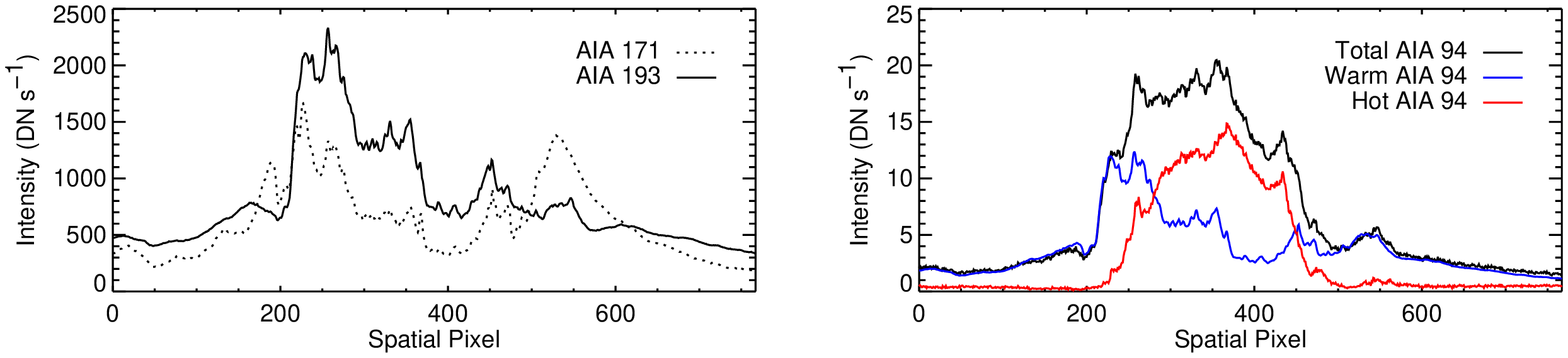}}
\caption{An example calculation of the \ion{Fe}{18} 93.92\,\AA\ intensity. The 171\,\AA\
  and 193\,\AA\ images are combined and scaled to estimate the warm contribution to the
  94\,\AA\ channel. The warm emission is then subtracted from the observed 94\,\AA\
  channel image to yield an estimate of the \ion{Fe}{18} 93.92\,\AA\ intensity. The bottom
  panels show the intensity as a function of position across the image. The intensities
  are averaged along the region indicated by the dotted lines. These data are from 2011
  November 8 near 16:34 UT.}
\label{fig:test1}
\end{figure*}

A number of previous studies have suggested that emission measure analysis is of little
utility since the inversion of equation~\ref{eq:ints} is ill-posed
\citep[e.g.,][]{craig1976,judge1997}. It is clear, however, that the general properties of
active region temperature structure can be determined from the available data. We can, for
example, safely conclude that the emission measure near 4\,MK is approximately 100 times
larger than the emission measure near 1\,MK in many of these active regions. This result
is evident in all of the Monte Carlo runs and in many different active regions, so is
robust against perturbations in the observed intensities. It is also clear that the
detailed structure of the emission measure distributions is much more difficult to
determine with confidence. Small changes in the parameters used in the inversion can lead
to different results \citep[e.g.,][]{landi2012}. If we run the MCMC code with a smaller
temperature binning, for example, we obtain distributions with much more structure. In the
example shown in Figure~\ref{fig:dem_alt}, the general trend is preserved, but the
emission measure distribution appears to break up into a series of nearly isothermal
components (see \citealt{landi2008} for a similar result). Understanding the detailed
structure of the emission measure distribution will require more detailed mathematical
analysis. At present, however, developing models of the coronal heating process which make
predictions comparable to the observations described here is likely to lead to the most
rapid progress on this long standing problem in solar physics.

%% ------------------------------------------------------------------------------------------
%% --- ACKNOWLEDGMENTS ----------------------------------------------------------------------
%% ------------------------------------------------------------------------------------------

\acknowledgments This research was supported by NASA. Hinode is a Japanese mission
developed and launched by ISAS/JAXA, with NAOJ as domestic partner and NASA and STFC (UK)
as international partners. It is operated by these agencies in co-operation with ESA and
NSC (Norway). HPW benefited greatly from discussions at a International Space Science
Institute meeting on coronal heating lead by Steve Bradshaw and Helen Mason. 

\appendix
\section{An Empirical Correction to the AIA 94 Channel}\label{sec:appendixA}

The AIA 94\,\AA\ channel is contaminated by ``warm'' emission formed at temperatures much
less than the 7.0\,MK temperature characteristic of \ion{Fe}{18}. To illustrate this we
have taken one hour of AIA observations (2010 March 22 12--13UT) and computed
time-averaged images from all of the available data. These data were chosen because they
contained a large bright point in addition to the quiet Sun and show a relatively large
range of intensities. The bright point, however, is unlikely to contain any \ion{Fe}{18},
which would complicate the analysis. The averaging naturally leads to some smearing of the
images but is necessary to improve the signal to noise. The averaged images for 5
wavelengths are shown in Figure~\ref{fig:qs1}.

Inspection of these images suggests that the warm emission is closest to 193 in
morphology. Note the strong contrast between the bright point and the quiet Sun, for
example. A detailed examination of the loops around the bright point indicates that there
is also a contribution from cooler emission similar in temperature to 171\,\AA. See
\citep{testa2012} for a discussion of stellar observations of the this wavelength
range. To estimate the intensities in the 94\,\AA\ channel we consider a polynomial fit to
an mixture of 171 and 193\,\AA\ images
\begin{equation}
I_{94warm} = 0.39\sum_{i=0}^4 a_i \left[\frac{fI_{171}+(1-f)I_{193}}{116.54}\right]^i,
\end{equation}
where the scaling factors derived from the median intensities (116.54 and 0.39) have been
introduced for convenience. We have determined that for $f=0.31$ the estimated intensities
are closest to what is observed. For this value of $f$ the coefficients to the polynomial
fit are $-7.31\times10^{-2}$, $9.75\times10^{-1}$, $9.90\times10^{-2}$, and
$-2.84\times10^{-3}$. Since there is very little data for very high intensities in these
data we limit the value of the composite 171/193\,\AA\ intensity to 30 in using the
polynomial fit. The observed and estimated 94\,\AA\ intensities for these data are shown
in Figure~\ref{fig:qs1}. 

An example set of images is shown in Figure~\ref{fig:test1}. These data were considered by
\citet{teriaca2012} and compared with spectroscopic observations of \ion{Fe}{18}
974.86\,\AA. Note that this procedure will not work during a flare since \ion{Fe}{24}
192.04\,\AA\ is likely to contribute to the 193\,\AA\ channel. This approach will also run
into problems for very bright 1\,MK emission, such as is found in the moss.

A similar method for isolating the \ion{Fe}{18} in the AIA 94\,\AA\ channel was
considered by \cite{reale2011}. They used only 171\,\AA, however, which does not
approximate the contaminating emission as well as a combination of 171\,\AA\ and
193\,\AA. An innovative technique for visualizing the relative contributions of active
region emission at various temperatures, including the very high temperature
\ion{Fe}{18} emission, has been presented by \cite{testa2012}.

%% ------------------------------------------------------------------------------------------
%% --- REFERENCES ---------------------------------------------------------------------------
%% ------------------------------------------------------------------------------------------

\bibliography{apj}

\begin{thebibliography}{}

\bibitem[\protect\citeauthoryear{{Aschwanden} \& {Nightingale}}{{Aschwanden} \&
  {Nightingale}}{2005}]{aschwanden2005}
{Aschwanden}, M.~J.,  \& {Nightingale}, R.~W. 2005, \apj, 633, 499

\bibitem[\protect\citeauthoryear{{Asgari-Targhi} \& {van
  Ballegooijen}}{{Asgari-Targhi} \& {van
  Ballegooijen}}{2012}]{asgari-targhi2012}
{Asgari-Targhi}, M.,  \& {van Ballegooijen}, A.~A. 2012, \apj, 746, 81

\bibitem[\protect\citeauthoryear{{Berger} et~al.}{{Berger}
  et~al.}{1999}]{berger1999}
{Berger}, T.~E., {de Pontieu}, B., {Fletcher}, L., {Schrijver}, C.~J.,
  {Tarbell}, T.~D.,  \& {Title}, A.~M. 1999, \solphys, 190, 409

\bibitem[\protect\citeauthoryear{{Cargill}}{{Cargill}}{1994}]{cargill1994}
{Cargill}, P.~J. 1994, \apj, 422, 381

\bibitem[\protect\citeauthoryear{{Cargill} \& {Klimchuk}}{{Cargill} \&
  {Klimchuk}}{2004}]{cargill2004}
{Cargill}, P.~J.,  \& {Klimchuk}, J.~A. 2004, \apj, 605, 911

\bibitem[\protect\citeauthoryear{{Craig} \& {Brown}}{{Craig} \&
  {Brown}}{1976}]{craig1976}
{Craig}, I.~J.~D.,  \& {Brown}, J.~C. 1976, \aap, 49, 239

\bibitem[\protect\citeauthoryear{{Culhane} et~al.}{{Culhane}
  et~al.}{2007}]{culhane2007}
{Culhane}, J.~L., et~al. 2007, \solphys, 243, 19

\bibitem[\protect\citeauthoryear{{De Rosa} et~al.}{{De Rosa}
  et~al.}{2009}]{derosa2009}
{De Rosa}, M.~L., et~al. 2009, \apj, 696, 1780

\bibitem[\protect\citeauthoryear{{Del Zanna}}{{Del Zanna}}{2012}]{delzanna2012}
{Del Zanna}, G. 2012, \aap, 537, A38

\bibitem[\protect\citeauthoryear{{Dere} et~al.}{{Dere} et~al.}{2009}]{dere2009}
{Dere}, K.~P., {Landi}, E., {Young}, P.~R., {Del Zanna}, G., {Landini}, M.,  \&
  {Mason}, H.~E. 2009, \aap, 498, 915

\bibitem[\protect\citeauthoryear{{Desai} et~al.}{{Desai}
  et~al.}{2005}]{desai2005}
{Desai}, P., et~al. 2005, \apjl, 625, L59

\bibitem[\protect\citeauthoryear{{Feldman} et~al.}{{Feldman}
  et~al.}{1992}]{feldman1992}
{Feldman}, U., {Mandelbaum}, P., {Seely}, J.~F., {Doschek}, G.~A.,  \&
  {Gursky}, H. 1992, \apjs, 81, 387

\bibitem[\protect\citeauthoryear{{Fludra} \& {Ireland}}{{Fludra} \&
  {Ireland}}{2008}]{fludra2008}
{Fludra}, A.,  \& {Ireland}, J. 2008, \aap, 483, 609

\bibitem[\protect\citeauthoryear{{Judge}, {Hubeny}, \& {Brown}}{{Judge}
  et~al.}{1997}]{judge1997}
{Judge}, P.~G., {Hubeny}, V.,  \& {Brown}, J.~C. 1997, \apj, 475, 275

\bibitem[\protect\citeauthoryear{{Kashyap} \& {Drake}}{{Kashyap} \&
  {Drake}}{1998}]{kashyap1998}
{Kashyap}, V.,  \& {Drake}, J.~J. 1998, \apj, 503, 450

\bibitem[\protect\citeauthoryear{{Kashyap} \& {Drake}}{{Kashyap} \&
  {Drake}}{2000}]{kashyap2000}
{Kashyap}, V.,  \& {Drake}, J.~J. 2000, Bulletin of the Astronomical Society of
  India, 28, 475

\bibitem[\protect\citeauthoryear{{Klimchuk} \& {Cargill}}{{Klimchuk} \&
  {Cargill}}{2001}]{klimchuk2001}
{Klimchuk}, J.~A.,  \& {Cargill}, P.~J. 2001, \apj, 553, 440

\bibitem[\protect\citeauthoryear{{Ko} et~al.}{{Ko} et~al.}{2009}]{ko2009}
{Ko}, Y., {Doschek}, G.~A., {Warren}, H.~P.,  \& {Young}, P.~R. 2009, \apj,
  697, 1956

\bibitem[\protect\citeauthoryear{{Korendyke} et~al.}{{Korendyke}
  et~al.}{2006}]{korendyke2006}
{Korendyke}, C.~M., et~al. 2006, \ao, 45, 8674

\bibitem[\protect\citeauthoryear{{Landi} \& {Feldman}}{{Landi} \&
  {Feldman}}{2008}]{landi2008}
{Landi}, E.,  \& {Feldman}, U. 2008, \apj, 672, 674

\bibitem[\protect\citeauthoryear{{Landi}, {Reale}, \& {Testa}}{{Landi}
  et~al.}{2012}]{landi2012}
{Landi}, E., {Reale}, F.,  \& {Testa}, P. 2012, \aap, 538, A111

\bibitem[\protect\citeauthoryear{{Lemen} et~al.}{{Lemen}
  et~al.}{2012}]{lemen2012}
{Lemen}, J.~R., et~al. 2012, \solphys, 275, 17

\bibitem[\protect\citeauthoryear{{Liu} et~al.}{{Liu} et~al.}{2012}]{liu2012}
{Liu}, Y., et~al. 2012, solphys, in press

\bibitem[\protect\citeauthoryear{{Mulu-Moore} et~al.}{{Mulu-Moore}
  et~al.}{2011}]{mulumoore2011}
{Mulu-Moore}, F.~M., {Winebarger}, A.~R., {Warren}, H.~P.,  \& {Aschwanden},
  M.~J. 2011, \apj, 733, 59

\bibitem[\protect\citeauthoryear{{O'Dwyer} et~al.}{{O'Dwyer}
  et~al.}{2010}]{odwyer2010}
{O'Dwyer}, B., {Del Zanna}, G., {Mason}, H.~E., {Weber}, M.~A.,  \& {Tripathi},
  D. 2010, \aap, 521, A21

\bibitem[\protect\citeauthoryear{{Parker}}{{Parker}}{1988}]{parker1988}
{Parker}, E.~N. 1988, \apj, 330, 474

\bibitem[\protect\citeauthoryear{{Peter}, {Bingert}, \& {Kamio}}{{Peter}
  et~al.}{2012}]{peter2012}
{Peter}, H., {Bingert}, S.,  \& {Kamio}, S. 2012, \aap, 537, A152

\bibitem[\protect\citeauthoryear{{Reale} et~al.}{{Reale}
  et~al.}{2011}]{reale2011}
{Reale}, F., {Guarrasi}, M., {Testa}, P., {DeLuca}, E.~E., {Peres}, G.,  \&
  {Golub}, L. 2011, \apjl, 736, L16

\bibitem[\protect\citeauthoryear{{Scherrer} et~al.}{{Scherrer}
  et~al.}{1995}]{scherrer1995}
{Scherrer}, P.~H., et~al. 1995, \solphys, 162, 129

\bibitem[\protect\citeauthoryear{{Scherrer} et~al.}{{Scherrer}
  et~al.}{2012}]{scherrer2012}
{Scherrer}, P.~H., et~al. 2012, \solphys, 275, 207

\bibitem[\protect\citeauthoryear{{Schrijver}}{{Schrijver}}{1987}]{schrijver1987}
{Schrijver}, C.~J. 1987, \aap, 180, 241

\bibitem[\protect\citeauthoryear{{Serio} et~al.}{{Serio}
  et~al.}{1981}]{serio1981}
{Serio}, S., {Peres}, G., {Vaiana}, G.~S., {Golub}, L.,  \& {Rosner}, R. 1981,
  \apj, 243, 288

\bibitem[\protect\citeauthoryear{{Teriaca}, {Warren}, \& {Curdt}}{{Teriaca}
  et~al.}{2012}]{teriaca2012}
{Teriaca}, L., {Warren}, H.~P.,  \& {Curdt}, W. 2012, \apj, in preparation

\bibitem[\protect\citeauthoryear{{Testa} \& {Reale}}{{Testa} \&
  {Reale}}{2012}]{testa2012}
{Testa}, P.,  \& {Reale}, F. 2012, \apjl, 750, L10

\bibitem[\protect\citeauthoryear{{Tripathi}, {Klimchuk}, \& {Mason}}{{Tripathi}
  et~al.}{2011}]{tripathi2011}
{Tripathi}, D., {Klimchuk}, J.~A.,  \& {Mason}, H.~E. 2011, \apj, 740, 111

\bibitem[\protect\citeauthoryear{{Tripathi} et~al.}{{Tripathi}
  et~al.}{2009}]{tripathi2009}
{Tripathi}, D., {Mason}, H.~E., {Dwivedi}, B.~N., {del Zanna}, G.,  \& {Young},
  P.~R. 2009, \apj, 694, 1256

\bibitem[\protect\citeauthoryear{{Tripathi}, {Mason}, \& {Klimchuk}}{{Tripathi}
  et~al.}{2010}]{tripathi2010}
{Tripathi}, D., {Mason}, H.~E.,  \& {Klimchuk}, J.~A. 2010, \apj, 723, 713

\bibitem[\protect\citeauthoryear{{Ugarte-Urra}, {Warren}, \&
  {Brooks}}{{Ugarte-Urra} et~al.}{2009}]{ugarteurra2009}
{Ugarte-Urra}, I., {Warren}, H.~P.,  \& {Brooks}, D.~H. 2009, \apj, 695, 642

\bibitem[\protect\citeauthoryear{{van Ballegooijen} et~al.}{{van Ballegooijen}
  et~al.}{2011}]{vanballegooijen2011}
{van Ballegooijen}, A.~A., {Asgari-Targhi}, M., {Cranmer}, S.~R.,  \& {DeLuca},
  E.~E. 2011, \apj, 736, 3

\bibitem[\protect\citeauthoryear{{Viall} \& {Klimchuk}}{{Viall} \&
  {Klimchuk}}{2011}]{viall2011}
{Viall}, N.~M.,  \& {Klimchuk}, J.~A. 2011, \apj, 738, 24

\bibitem[\protect\citeauthoryear{{Warren}, {Brooks}, \& {Winebarger}}{{Warren}
  et~al.}{2011}]{warren2011}
{Warren}, H.~P., {Brooks}, D.~H.,  \& {Winebarger}, A.~R. 2011, \apj, 734, 90

\bibitem[\protect\citeauthoryear{{Warren}, {Feldman}, \& {Brown}}{{Warren}
  et~al.}{2008}]{warren2008b}
{Warren}, H.~P., {Feldman}, U.,  \& {Brown}, C.~M. 2008, \apj, 685, 1277

\bibitem[\protect\citeauthoryear{{Warren} et~al.}{{Warren}
  et~al.}{2008}]{warren2008}
{Warren}, H.~P., {Ugarte-Urra}, I., {Doschek}, G.~A., {Brooks}, D.~H.,  \&
  {Williams}, D.~R. 2008, \apjl, 686, L131

\bibitem[\protect\citeauthoryear{{Warren} \& {Winebarger}}{{Warren} \&
  {Winebarger}}{2006}]{warren2006}
{Warren}, H.~P.,  \& {Winebarger}, A.~R. 2006, \apj, 645, 711

\bibitem[\protect\citeauthoryear{{Winebarger} et~al.}{{Winebarger}
  et~al.}{2011}]{winebarger2011}
{Winebarger}, A.~R., {Schmelz}, J.~T., {Warren}, H.~P., {Saar}, S.~H.,  \&
  {Kashyap}, V.~L. 2011, \apj, 740, 2

\bibitem[\protect\citeauthoryear{{Winebarger}, {Warren}, \&
  {Seaton}}{{Winebarger} et~al.}{2003}]{winebarger2003}
{Winebarger}, A.~R., {Warren}, H.~P.,  \& {Seaton}, D.~B. 2003, \apj, 593, 1164

\end{thebibliography}

\end{document}